\def\Teff{$T_{\rm eff}$} 
\def\Tex{$T_{\rm ex}$}
\newcommand{\Msun}{\ifmmode {M_{\odot}} \else ${M_{\odot}}$ \fi} 
\def\Macc{$\dot{M}_{\rm acc}$} 
\newcommand{\Rsun}{\ifmmode {R_{\odot}} \else ${R_{\odot}}$ \fi} 
\newcommand{\Lsun}{\ifmmode {L_{\odot}} \else ${L_{\odot}}$ \fi} 
\newcommand{\Lbol}{\ifmmode {L_{\rm bol}} \else ${L_{\rm bol}}$ \fi} 
\def\Lstar{$L_{\rm *}$}
\def\Mstar{$M_{\rm *}$}
\def\Rstar{$R_{\rm *}$}
\def\Lacc{$L_{\rm acc}$}

\def\Tbol{${T_{\rm bol}}$}

\def\Vmax{$\textit{v}_{max}$}

\newcommand{\FeI}{Fe\,{\footnotesize I}}
\newcommand{\FeII}{Fe\,{\footnotesize II}}

\newcommand{\MgI}{Mg\,{\footnotesize I}}

\newcommand{\kms}{\ifmmode \,\rm km\,s^{-1} \else $\,\rm km\,s^{-1}$ \fi}

\newcommand{\um}{$\mu$m}
\newcommand{\hmol}{H$_{2}$}
\newcommand{\yr}{\ifmmode {\rm yr^{-1}} \else yr${^{-1}}$ \fi} 
\newcommand{\Av}{$A_V$}

\newcommand\konkoly{Konkoly Observatory, Research Centre for Astronomy and Earth Sciences, E\"otv\"os Lor\'and Research Network (ELKH), Konkoly-Thege Mikl\'os \'ut 15-17, 1121 Budapest, Hungary}
\newcommand\csfk{CSFK, MTA Centre of Excellence, Konkoly-Thege Mikl\'os \'ut 15-17, 1121 Budapest, Hungary}
\newcommand\elte{ELTE E\"otv\"os Lor\'and University, Institute of Physics, P\'azm\'any P\'eter s\'et\'any 1/A, 1117 Budapest, Hungary}

\documentclass[arguments]{aastex631}
\shorttitle{Gaia19fct}
\shortauthors{Park et al.}
\graphicspath{{./}{figures/}}

\begin{document}

\title{Photometric and spectroscopic study of the EXor-like eruptive young star Gaia19fct}

\correspondingauthor{Sunkyung Park}
\email{sunkyung.park@csfk.org}

\author[0000-0003-4099-1171]{Sunkyung Park}
\affiliation{\konkoly{}}
\affiliation{\csfk{}}

\author[0000-0001-7157-6275]{\'Agnes K\'osp\'al}
\affiliation{\konkoly{}}
\affiliation{\csfk{}}
\affiliation{Max Planck Institute for Astronomy, Königstuhl 17, 69117 Heidelberg, Germany}
\affiliation{\elte{}}

\author[0000-0001-6015-646X]{P\'eter \'Abrah\'am}
\affiliation{\konkoly{}}
\affiliation{\csfk{}}
\affiliation{\elte{}}

\author[0000-0002-4283-2185]{Fernando Cruz-S\'aenz de Miera}
\affiliation{\konkoly{}}
\affiliation{\csfk{}}

\author[0000-0002-5261-6216]{Eleonora Fiorellino}
\affiliation{\konkoly{}}
\affiliation{\csfk{}}
\affiliation{INAF-Osservatorio Astronomico di Capodimonte, via Moiariello 16, 80131 Napoli, Italy}

\author[0000-0001-5018-3560]{Micha{\l} Siwak}
\affiliation{\konkoly{}}
\affiliation{\csfk{}}

\author[0000-0002-3632-1194]{Zs\'ofia Nagy}
\affiliation{\konkoly{}}
\affiliation{\csfk{}}

\author[0000-0002-7035-8513]{Teresa Giannini}
\affiliation{INAF-Osservatorio Astronomico di Roma, Via Frascati 33, 00078 Monte Porzio Catone, Italy}

\author[0000-0003-1604-2064]{Roberta Carini}
\affiliation{INAF-Osservatorio Astronomico di Roma, Via Frascati 33, 00078 Monte Porzio Catone, Italy}

\author[0000-0001-9830-3509]{Zs\'ofia Marianna Szab\'o}
\affiliation{\konkoly{}}
\affiliation{\csfk{}}
\affiliation{Max-Planck-Institute f\"{u}r Radioastronomie, Auf dem H\"{u}gel 69, 53121 Bonn, Germany}
\affiliation{Scottish Universities Physics Alliance (SUPA), School of Physics and Astronomy, University of St Andrews, North Haugh, St Andrews, KY16 9SS, UK}

\author[0000-0003-3119-2087]{Jeong-Eun Lee}
\affiliation{Department of Physics and Astronomy, Seoul National University, 1 Gwanak-ro, Gwanak-gu, Seoul 08826, Republic of Korea}

\author[0000-0003-0894-7824]{Jae-Joon Lee}
\affiliation{Korea Astronomy and Space Science Institute 776, Daedeok-daero, Yuseong-gu, Daejeon, 34055, Republic of Korea}

\author[0000-0001-8332-4227]{Fabrizio Vitali}
\affiliation{INAF-Osservatorio Astronomico di Roma, Via Frascati 33, 00078 Monte Porzio Catone, Italy}

\author[0000-0002-7538-5166]{M\'aria Kun}
\affiliation{\konkoly{}}
\affiliation{\csfk{}}

\author{Borb\'ala Cseh}
\affiliation{\konkoly{}}
\affiliation{\csfk{}}
\affiliation{MTA-ELTE Lend{\"u}let "Momentum" Milky Way Research Group, Hungary}
\author[0000-0002-8813-4884]{M\'at\'e Krezinger}
\affiliation{\konkoly{}}
\affiliation{\csfk{}}
\author[0000-0002-1792-546X]{Levente Kriskovics}
\affiliation{\konkoly{}}
\affiliation{\csfk{}}
\author{Andr\'as Ordasi}
\affiliation{\konkoly{}}
\affiliation{\csfk{}}
\author[0000-0001-5449-2467]{Andr\'as P\'al}
\affiliation{\konkoly{}}
\affiliation{\csfk{}}
\author[0000-0002-1698-605X]{R\'obert Szak\'ats}
\affiliation{\konkoly{}}
\affiliation{\csfk{}}
\author[0000-0002-6471-8607]{Kriszti\'an Vida}
\affiliation{\konkoly{}}
\affiliation{\csfk{}}
\author[0000-0001-8764-7832]{J\'ozsef Vink\'o}
\affiliation{\konkoly{}}
\affiliation{\csfk{}}



\begin{abstract}
Gaia19fct is one of the Gaia-alerted eruptive young stars that has undergone several brightening events. We conducted monitoring observations using multi-filter optical and near-infrared photometry, as well as near-infrared spectroscopy, to understand the physical properties of Gaia19fct and investigate whether it fits into the historically defined two classes.
We present the analyses of light curves, color variations, spectral lines, and CO modeling. The light curves show at least five brightening events since 2015, and the multi-filter color evolutions are mostly gray. The gray evolution indicates that bursts are triggered by mechanisms other than extinction. Our near-infrared spectra exhibit both absorption and emission lines and show time-variability throughout our observations. We found lower rotational velocity and lower temperature from the near-infrared atomic absorption lines than from the optical lines, suggesting that Gaia19fct has a Keplerian rotating disk. The CO overtone features show a superposition of absorption and emission components, which is unlike other young stellar objects. We modeled the CO lines, and the result suggests that the emission and absorption components are formed in different regions. We found that although Gaia19fct exhibits characteristics of both types of eruptive young stars, FU Orionis-type objects (FUors) and EX Lupi-type objects (EXors), it shows more similarity with EXors in general.
\end{abstract}


\section{Introduction} \label{sec:intro}
Most young stellar objects (YSOs) exhibit photometric variations in the optical and infrared on various timescales \citep{carpenter2001, megeath2012, cody2014, park2021}.
The photometric variability of YSOs can be caused by changing accretion rate, varying line-of-sight extinction, or rotating accretion hot or cold spots \citep{carpenter2001, megeath2012, kraus2016, fischer2022}. Among these photometric variability mechanisms, eruptive YSOs show the most dramatic change in brightness of about 3--6 magnitudes caused by an enhanced mass accretion rate from the disk to the central protostar. Such brightenings are thought to be observable evidence of episodic accretion and a possible solution to the protostellar luminosity problem \citep{kenyon1990, dunham2010}, which is a discrepancy between the luminosity expected by a standard accretion model \citep{shu1977} and the observed luminosities of YSOs.
Targets showing such a sudden brightness change are historically classified into two categories \citep[][and references therein]{fischer2022}: FU Orionis-type objects \citep[FUors;][]{herbig1977} and EX Lupi-type objects \citep[EXors;][]{herbig1989}.

FUors exhibit a large amplitude of outburst ($\Delta$V~$>$~4\,mag) and last for several decades to centuries. Spectroscopically, FUors show mostly absorption line profiles formed by a hotter disk midplane and have wavelength-dependent spectral types: F-G type in optical and K-M type in near-infrared (NIR) \citep{hartmann1996, audard2014, connelley2018, fischer2022}. 
EXors have recurrent outbursts with typically smaller amplitude ($\Delta$V~$= 1- 4$\,mag) that last for several months to years and have an emission line dominated spectrum formed by magnetospheric accretion funnels \citep{herbig2007, audard2014, hartmann2016, fischer2022}.

Recent studies \citep[][and references therein]{fischer2022} show many of the eruptive young stars have peculiar properties in photometry and spectroscopy, making it difficult to classify into only two classes. 
Therefore, a detailed study of each eruptive star is important to understand episodic accretion.
Thanks to the whole sky monitoring of the Gaia space mission and its Gaia Photometric Science Alerts Program\footnote{\url{http://gsaweb.ast.cam.ac.uk/alerts/home}} \citep{hodgkin2021}, which announces targets that show an abrupt brightening or fading, the number of eruptive young stars has been increasing. 
These discoveries have been classified as FUors (Gaia17bpi \citep{hillenbrand2018}, Gaia18dvy \citep{szegedi-elek2020}), EXors (Gaia18dvz \citep{hodapp2019}, Gaia20eae \citep{ghosh2022, csm2022}), and in between (Gaia19ajj \citep{hillenbrand2019_gaia19ajj}, Gaia19bey \citep{hodapp2020}).

Gaia19fct (also known as iPTF~15afq) was discovered on 2015 March 13 \citep[$\Delta$R~$\sim$~2.5\,mag;][]{miller2015}, and the outbursting optical spectrum showed rich emission lines. Based on the amplitude of its brightening and rich emission lines, Gaia19fct was suggested as a new EXor \citep{miller2015}.
Repeated brightenings in 2018 and 2019 were reported by \citet{hillenbrand2019}. The brightening in 2019 ($\Delta$G~$>$~4\,mag) was reported by the Gaia Photometric Science Alerts Program on 2019 November 14. \citet{giannini2022} presented optical and NIR spectra of Gaia19fct. The quiescent spectrum is dominated by emission lines, while the outburst spectrum is dominated by absorption lines \citep{hillenbrand2019, giannini2022}.
The overall amplitudes, time scales of brightness events, and spectroscopic properties of Gaia19fct are more similar to EXors.

In this study, we present the results of our monitoring program for this target, which started in 2016 September and included photometry and spectroscopy in optical and NIR. We characterize the physical properties of Gaia19fct by analyzing the results of our observational campaigns.
Our observations, data reduction, and the used public domain data are described in Section~\ref{sec:observation}. We discuss the environment of Gaia19fct in Section~\ref{sec:gaia19fct}. 
The classification of Gaia19fct and the analyses of photometry and spectroscopy are presented in Section~\ref{sec:results}. In Section~\ref{sec:discussion}, we discuss and compare our results with previous studies. Finally, we summarize our results and findings in Section~\ref{sec:conclusion}.

\section{Observation} \label{sec:observation}
\subsection{Photometry}
\subsubsection{Optical Photometry}
We have been monitoring Gaia19fct since 2020 October with the 80\,cm Ritchey-Chr\'etien (RC80) telescope at the Piszk\'estet\H{o} mountain station of Konkoly Observatory in Hungary. The telescope is equipped with an FLI PL230 CCD camera, 0$\farcs$55 pixel scale, 18$\farcm$8$\times$18$\farcm$8 field of view, and Johnson $V$ and Sloan $g'r'i'$ filters. After standard bias, dark, and flat-field correction, we co-added the three images per filter per night. We obtained aperture photometry in the co-added images for Gaia19fct and for 40 comparison stars within 10$'$ of the science target. We used an aperture radius of 5 pixels (2$\farcs$75) and sky annulus between 20 and 40 pixels (11$''$ and 22$''$). We used the APASS9 magnitudes \citep{henden2015} of the comparison stars for photometric calibration by fitting a linear color term. The results are listed in \autoref{tbl_rc80_rem_photometry}.

\begin{deluxetable*}{cccccccccc}
\tabletypesize{\scriptsize} 
\tablecaption{Our Optical and NIR photometry \label{tbl_rc80_rem_photometry}}
\tablewidth{0pt}
\tablehead{
\colhead{JD} & \colhead{Date [UT]} & \colhead{$g'$} & \colhead{$V$} & \colhead{$r'$} & \colhead{$i'$}  & \colhead{$J$} & \colhead{$H$}& \colhead{$K_s$} & \colhead{Instrument}
}
\startdata
2457649.00	&	2016-09-17	&	\dots			&	\dots			&	\dots			&	\dots			&	13.32	$\pm$	0.08	&	11.85	$\pm$	0.11	&	10.61	$\pm$	0.10	&	REMIR	\\
2457664.00	&	2016-10-02	&	\dots			&	\dots			&	\dots			&	\dots			&	14.56	$\pm$	0.13	&	12.83	$\pm$	0.10	&	11.31	$\pm$	0.10	&	REMIR	\\
2457682.00	&	2016-10-20	&	\dots			&	\dots			&	\dots			&	\dots			&	14.47	$\pm$	0.15	&	12.75	$\pm$	0.11	&	11.35	$\pm$	0.21	&	REMIR	\\
2457697.00	&	2016-11-04	&	\dots			&	\dots			&	\dots			&	\dots			&	14.45	$\pm$	0.13	&	12.57	$\pm$	0.09	&	10.90	$\pm$	0.08	&	REMIR	\\
2457713.00	&	2016-11-20	&	\dots			&	\dots			&	\dots			&	\dots			&	13.80	$\pm$	0.10	&	12.17	$\pm$	0.09	&	11.50	$\pm$	0.03	&	REMIR	\\
2457729.00	&	2016-12-06	&	\dots			&	\dots			&	\dots			&	\dots			&	13.20	$\pm$	0.14	&	11.78	$\pm$	0.22	&	10.37	$\pm$	0.15	&	REMIR	\\
2457744.00	&	2016-12-21	&	\dots			&	\dots			&	\dots			&	\dots			&	\dots			&	11.73	$\pm$	0.12	&	10.55	$\pm$	0.14	&	REMIR	\\
2457759.00	&	2017-01-05	&	\dots			&	\dots			&	\dots			&	\dots			&	13.27	$\pm$	0.17	&	12.08	$\pm$	0.08	&	10.60	$\pm$	0.18	&	REMIR	\\
2457774.00	&	2017-01-20	&	\dots			&	\dots			&	\dots			&	\dots			&	13.25	$\pm$	0.12	&	11.81	$\pm$	0.24	&	10.76	$\pm$	0.21	&	REMIR	\\
2457792.00	&	2017-02-07	&	\dots			&	\dots			&	\dots			&	\dots			&	13.66	$\pm$	0.13	&	12.14	$\pm$	0.22	&	11.66	$\pm$	0.09	&	REMIR	\\
2457815.00	&	2017-03-02	&	\dots			&	\dots			&	\dots			&	\dots			&	14.03	$\pm$	0.18	&	12.47	$\pm$	0.18	&	11.15	$\pm$	0.14	&	REMIR	\\
2457891.00	&	2017-05-17	&	\dots			&	\dots			&	\dots			&	\dots			&	14.08	$\pm$	0.11	&	12.72	$\pm$	0.05	&	11.63	$\pm$	0.06	&	REMIR	\\
2457907.00	&	2017-06-02	&	\dots			&	\dots			&	\dots			&	\dots			&	14.10	$\pm$	0.20	&	\dots			&	\dots			&	REMIR	\\
2457908.00	&	2017-06-03	&	\dots			&	\dots			&	\dots			&	\dots			&	\dots			&	12.86	$\pm$	0.10	&	11.83	$\pm$	0.11	&	REMIR	\\
2457991.00	&	2017-08-25	&	\dots			&	\dots			&	\dots			&	\dots			&	14.13	$\pm$	0.11	&	12.68	$\pm$	0.07	&	12.05	$\pm$	0.13	&	REMIR	\\
2458008.00	&	2017-09-11	&	\dots			&	\dots			&	\dots			&	\dots			&	14.08	$\pm$	0.14	&	12.83	$\pm$	0.15	&	11.96	$\pm$	0.12	&	REMIR	\\
2458187.00	&	2018-03-09	&	\dots			&	\dots			&	\dots			&	\dots			&	14.29	$\pm$	0.15	&	12.99	$\pm$	0.19	&	12.10	$\pm$	0.10	&	REMIR	\\
2458202.00	&	2018-03-24	&	\dots			&	\dots			&	\dots			&	\dots			&	13.90	$\pm$	0.12	&	12.47	$\pm$	0.14	&	11.45	$\pm$	0.20	&	REMIR	\\
2458217.00	&	2018-04-08	&	\dots			&	\dots			&	\dots			&	\dots			&	11.71	$\pm$	0.04	&	10.39	$\pm$	0.03	&	9.25	$\pm$	0.02	&	REMIR	\\
2458236.00	&	2018-04-27	&	\dots			&	\dots			&	\dots			&	\dots			&	11.79	$\pm$	0.10	&	10.37	$\pm$	0.15	&	9.30	$\pm$	0.18	&	REMIR	\\
2458251.00	&	2018-05-12	&	\dots			&	\dots			&	\dots			&	\dots			&	11.99	$\pm$	0.07	&	10.62	$\pm$	0.05	&	9.50	$\pm$	0.08	&	REMIR	\\
2458346.00	&	2018-08-15	&	\dots			&	\dots			&	\dots			&	\dots			&	14.39	$\pm$	0.15	&	\dots			&	\dots			&	REMIR	\\
2458350.00	&	2018-08-19	&	\dots			&	\dots			&	\dots			&	\dots			&	14.28	$\pm$	0.18	&	12.72	$\pm$	0.11	&	10.98	$\pm$	0.03	&	REMIR	\\
2458351.00	&	2018-08-20	&	\dots			&	\dots			&	\dots			&	\dots			&	14.48	$\pm$	0.16	&	12.83	$\pm$	0.08	&	11.13	$\pm$	0.10	&	REMIR	\\
2458369.00	&	2018-09-07	&	\dots			&	\dots			&	\dots			&	\dots			&	13.98	$\pm$	0.12	&	12.32	$\pm$	0.08	&	10.93	$\pm$	0.07	&	REMIR	\\
2458385.00	&	2018-09-23	&	\dots			&	\dots			&	\dots			&	\dots			&	13.76	$\pm$	0.18	&	\dots			&	10.88	$\pm$	0.05	&	REMIR	\\
2459145.60	&	2020-10-23	&	19.72	$\pm$	0.24	&	18.61	$\pm$	0.16	&	17.54	$\pm$	0.08	&	16.42	$\pm$	0.04	&	\dots			&	\dots			&	\dots			&	RC80	\\
2459154.66	&	2020-11-01	&	19.58	$\pm$	0.25	&	18.35	$\pm$	0.10	&	17.87	$\pm$	0.07	&	16.55	$\pm$	0.03	&	\dots			&	\dots			&	\dots			&	RC80	\\
2459161.59	&	2020-11-08	&	19.86	$\pm$	0.15	&	19.06	$\pm$	0.13	&	17.77	$\pm$	0.06	&	16.75	$\pm$	0.05	&	\dots			&	\dots			&	\dots			&	RC80	\\
2459175.62	&	2020-11-22	&	20.11	$\pm$	0.22	&	19.27	$\pm$	0.19	&	17.88	$\pm$	0.08	&	16.73	$\pm$	0.05	&	\dots			&	\dots			&	\dots			&	RC80	\\
2459491.61	&	2021-10-04	&	18.70	$\pm$	0.06	&	17.77	$\pm$	0.04	&	16.59	$\pm$	0.03	&	15.51	$\pm$	0.03	&	\dots			&	\dots			&	\dots			&	RC80	\\
2459524.61	&	2021-11-06	&	\dots			&	\dots			&	16.95	$\pm$	0.05	&	15.82	$\pm$	0.05	&	\dots			&	\dots			&	\dots			&	RC80	\\
2459535.54	&	2021-11-17	&	20.36	$\pm$	0.39	&	19.24	$\pm$	0.23	&	18.07	$\pm$	0.12	&	16.50	$\pm$	0.07	&	\dots			&	\dots			&	\dots			&	RC80	\\
2459542.58	&	2021-11-24	&	21.50	$\pm$	0.38	&	20.07	$\pm$	0.19	&	18.66	$\pm$	0.08	&	17.27	$\pm$	0.05	&	\dots			&	\dots			&	\dots			&	RC80	\\
2459546.00	&	2021-11-27	&	\dots			&	\dots			&	\dots			&	\dots			&	13.40	$\pm$	0.12	&	11.73	$\pm$	0.05	&	10.20	$\pm$	0.14	&	REMIR	\\
2459548.54	&	2021-11-30	&	20.93	$\pm$	0.27	&	19.72	$\pm$	0.16	&	18.86	$\pm$	0.09	&	17.35	$\pm$	0.04	&	\dots			&	\dots			&	\dots			&	RC80	\\
2459552.53	&	2021-12-04	&	20.66	$\pm$	0.13	&	19.77	$\pm$	0.13	&	18.61	$\pm$	0.05	&	17.15	$\pm$	0.04	&	\dots			&	\dots			&	\dots			&	RC80	\\
2459556.53	&	2021-12-08	&	20.76	$\pm$	0.15	&	20.02	$\pm$	0.15	&	18.48	$\pm$	0.06	&	17.16	$\pm$	0.05	&	\dots			&	\dots			&	\dots			&	RC80	\\
2459562.49	&	2021-12-13	&	20.73	$\pm$	0.16	&	19.61	$\pm$	0.10	&	18.37	$\pm$	0.05	&	16.84	$\pm$	0.04	&	\dots			&	\dots			&	\dots			&	RC80	\\
2459566.48	&	2021-12-17	&	\dots			&	20.60	$\pm$	0.60	&	18.44	$\pm$	0.11	&	17.09	$\pm$	0.08	&	\dots			&	\dots			&	\dots			&	RC80	\\
2459574.00	&	2021-12-25	&	\dots			&	\dots			&	\dots			&	\dots			&	13.04	$\pm$	0.14	&	11.43	$\pm$	0.13	&	10.39	$\pm$	0.26	&	REMIR	\\
2459581.56	&	2022-01-02	&	21.26	$\pm$	0.25	&	19.64	$\pm$	0.12	&	18.30	$\pm$	0.06	&	17.11	$\pm$	0.05	&	\dots			&	\dots			&	\dots			&	RC80	\\
2459590.00	&	2022-01-10	&	\dots			&	\dots			&	\dots			&	\dots			&	13.32	$\pm$	0.15	&	11.64	$\pm$	0.15	&	10.68	$\pm$	0.07	&	REMIR	\\
2459607.00	&	2022-01-27	&	\dots			&	\dots			&	\dots			&	\dots			&	13.29	$\pm$	0.19	&	11.92	$\pm$	0.06	&	10.63	$\pm$	0.16	&	REMIR	\\
2459614.28	&	2022-02-03	&	21.79	$\pm$	1.19	&	19.61	$\pm$	0.33	&	18.39	$\pm$	0.20	&	17.41	$\pm$	0.08	&	\dots			&	\dots			&	\dots			&	RC80	\\
2459622.00	&	2022-02-11	&	\dots			&	\dots			&	\dots			&	\dots			&	13.62	$\pm$	0.16	&	12.25	$\pm$	0.24	&	10.95	$\pm$	0.11	&	REMIR	\\
2459656.41	&	2022-03-17	&	\dots			&	\dots			&	\dots			&	\dots			&	\dots			&	11.77	$\pm$	0.04	&	10.48	$\pm$	0.01	&	NOTCam	\\
2459668.00	&	2022-03-29	&	\dots			&	\dots			&	\dots			&	\dots			&	13.20	$\pm$	0.12	&	11.92	$\pm$	0.06	&	\dots			&	REMIR	\\
2459857.65  &   2022-10-05  &   \dots           &   \dots           &   \dots           &  18.20 $\pm$   0.13 &   \dots &   \dots &   \dots &    RC80 \\
\enddata
\end{deluxetable*}

\subsubsection{Near-infrared Photometry}
We obtained photometric images on 2022 March 17 (Program ID: 65-111; PI: Cruz-S\'aenz de Miera) using the $H$ and $K_s$ filters of the NOTCam installed at the 2.56\,m Nordic Optical Telescope (NOT) located at Roque de los Muchachos Observatory (La Palma, Canary Islands, Spain). For each filter, we used a 5-point dither pattern with offsets of 60$''$.
The exposure time was 4 seconds in both filters. 
We constructed sky images by computing the median of the individual frames per filter, and subtracted it from the original frames (in the $K_s$ band, only 4 frames were used due to an instrumental artifact). 
Flat-field measurements were obtained by the observatory on the same night. For photometric calibration, we used all 2MASS sources detected in the frames brighter than 13.8\,mag ($H$ band) or 13.0\,mag ($K_s$ band), and have good quality flag of `A'. We determined the calibration factor (the difference between the 2MASS and instrumental magnitudes) by averaging 5$-$17 stars, and applied this factor to Gaia19fct. The influence of intrinsically variable comparison stars was minimized by using an outlier resistant averaging method.
We performed photometry on the individual frames and averaged the results, which also provided formal uncertainties of the final photometric values as the standard deviation of the individual photometric values.

The $JHK_{s}$ photometric monitoring were carried out between 2016 September and 2022 March with the 60-cm robotic Rapid Eye Mount (REM) telescope \citep{zerbi2001}, an Italian INAF facility located in La Silla (Chile), hosted by ESO. The NIR images were obtained with the infrared imaging camera REMIR \citep{vitali2003, vitali2006} with broadband $J$, $H$, and $K_{s}$ filters. All the observations were obtained by dithering the field of view around the pointed position, with a total exposure time of 15 seconds in each band. The raw frames were reduced using the Riace semi-automatic IRAF procedure (Vitali et al., in preparation), which uses the 2MASS catalog to define the average zero point for the photometric calibration, and performs aperture photometry to extract the magnitude of the source in each band. Observing results are listed in \autoref{tbl_rc80_rem_photometry}.

a\subsubsection{Public domain data}
To get an insight into the historical light variations of Gaia19fct in optical bands, we downloaded all individual Pan-STARRS \citep{chambers2016} images containing our target and analyzed them with {\sc DAOPHOT} procedures available within \texttt{astro-idl}. Due to the nearby companion (see more details in Section~\ref{sec:companion}), we used a small 5\,pixel aperture, which equals 2$\farcs$25 in the sky. We stress that this approach may occasionally slightly decrease (by 20\%) the total flux obtained from images taken during poor seeing conditions. Fortunately, these losses are small enough and isolated in time and do not affect our general conclusions regarding the large-scale historical variability. 
The magnitudes were calibrated to the standard system employing the nightly zero points provided in the fits headers. We checked that the $gri$ bands magnitudes of the three nearby standard stars from the APASS9 \citep{henden2015} catalog are in good agreement ($\pm$0.07\,mag) with our determinations. The results are listed in Table~\ref{tbl_ps1_photometry}.

Furthermore, we gathered public domain photometry data from various sources to complement our monitoring data. We used Gaia $G$-band photometry from the Gaia Photometric Science Alerts database and Zwicky Transient Facility \citep[ZTF;][]{masci2019} DR13 $g$- and $r$-band photometry from the ZTF archive. We used the ZTF data with ``catflags=0", i.e., perfectly clean extracted, to filter out bad-quality images. We also collected $VRI$ bands data from Hunting Outbursting Young Stars with the Centre of Astrophysics and Planetary Science (HOYS-CAPS\footnote{\url{http://astro.kent.ac.uk/HOYS-CAPS/}}) citizen science project run by the University of Kent \citep{froebrich2018, evitts2020}. We used the photometric data available from their websites for the Gaia, ZTF, and HOYS-CAPS surveys; therefore, instead of providing their values in the table, we plotted them in \autoref{fig:light_curve} and \autoref{fig:cmd}.

We downloaded 3.4$\,\mu$m (W1) and 4.6$\,\mu$m (W2) photometry for Gaia19fct from the catalogs of the WISE Cryogenic Survey, NEOWISE Post-Cryo Survey, and the NEOWISE Reactivation mission \citep{wright2010, mainzer2011, mainzer2014}. We used the AllWISE Multiepoch Photometry Table and the NEOWISE-R Single Exposure (L1b) Source Table available at NASA/IPAC Infrared Science Archive. After filtering for bad-quality data (using the qi\_fact, saa\_sep, and moon\_masked flags, for more details, see Section~II.3.a~of the NEOWISE Explanatory Supplement), we applied saturation correction using the appropriate correction curves provided by NEOWISE Explanatory Supplement (Section~II.1.c.iv.a). Finally, we calculated seasonal averages, as WISE typically scanned the sky twice per year. \autoref{tbl_wise_photometry} lists these averaged WISE magnitudes for Gaia19fct.

\begin{deluxetable}{ccccc}
\tabletypesize{\scriptsize} 
\tablecaption{Pan-STARRS Photometry of Gaia19fct and its companion\label{tbl_ps1_photometry}}
\tablewidth{0pt}
\tablehead{
\colhead{JD} & \colhead{Date [UT]} & \colhead{Gaia19fct} & \colhead{Companion} & \colhead{filter} 
}
\startdata
2455593.93	&	2011-02-01	&	22.20	$\pm$	0.20	&	21.99	$\pm$	0.17	&	g	\\
2455594.82	&	2011-02-02	&	22.12	$\pm$	0.18	&	22.14	$\pm$	0.18	&	g	\\
2455594.83	&	2011-02-02	&	22.54	$\pm$	0.27	&	22.52	$\pm$	0.26	&	g	\\
2455948.89	&	2012-01-22	&	22.20	$\pm$	0.24	&	22.09	$\pm$	0.22	&	g	\\
2455957.97	&	2012-01-31	&	22.04	$\pm$	0.17	&	22.63	$\pm$	0.27	&	g	\\
2456000.78	&	2012-03-14	&	22.36	$\pm$	0.22	&	22.01	$\pm$	0.17	&	g	\\
2456358.79	&	2013-03-07	&	22.83	$\pm$	0.37	&	22.06	$\pm$	0.19	&	g	\\
2456737.77	&	2014-03-21	&	21.89	$\pm$	0.17	&	22.03	$\pm$	0.20	&	g	\\
2456737.78	&	2014-03-21	&	22.39	$\pm$	0.28	&	22.50	$\pm$	0.31	&	g	\\
2455587.92	&	2011-01-26	&	20.63	$\pm$	0.06	&	21.06	$\pm$	0.07	&	r	\\
2455940.85	&	2012-01-14	&	19.40	$\pm$	0.02	&	20.93	$\pm$	0.06	&	r	\\
2455940.86	&	2012-01-14	&	19.42	$\pm$	0.02	&	20.97	$\pm$	0.07	&	r	\\
2456000.79	&	2012-03-14	&	20.14	$\pm$	0.04	&	20.92	$\pm$	0.07	&	r	\\
2456000.80	&	2012-03-14	&	20.24	$\pm$	0.05	&	20.98	$\pm$	0.07	&	r	\\
2456601.13	&	2013-11-04	&	19.60	$\pm$	0.03	&	21.05	$\pm$	0.08	&	r	\\
2456676.89	&	2014-01-19	&	19.98	$\pm$	0.05	&	20.89	$\pm$	0.10	&	r	\\
2456676.91	&	2014-01-19	&	20.05	$\pm$	0.06	&	20.97	$\pm$	0.10	&	r	\\
2456709.78	&	2014-02-21	&	20.96	$\pm$	0.09	&	\dots			&	r	\\
2455194.99	&	2009-12-29	&	19.42	$\pm$	0.04	&	19.49	$\pm$	0.04	&	i	\\
2455195.00	&	2009-12-29	&	19.46	$\pm$	0.04	&	19.46	$\pm$	0.04	&	i	\\
2455584.83	&	2011-01-23	&	19.82	$\pm$	0.06	&	19.45	$\pm$	0.04	&	i	\\
2455584.84	&	2011-01-23	&	19.74	$\pm$	0.05	&	19.43	$\pm$	0.04	&	i	\\
2455584.86	&	2011-01-23	&	19.78	$\pm$	0.05	&	19.39	$\pm$	0.04	&	i	\\
2455584.87	&	2011-01-23	&	19.67	$\pm$	0.05	&	19.43	$\pm$	0.04	&	i	\\
2455940.87	&	2012-01-14	&	17.83	$\pm$	0.01	&	\dots			&	i	\\
2455957.89	&	2012-01-31	&	18.95	$\pm$	0.02	&	19.35	$\pm$	0.02	&	i	\\
2455957.90	&	2012-01-31	&	18.93	$\pm$	0.02	&	19.33	$\pm$	0.02	&	i	\\
2456313.90	&	2013-01-21	&	18.59	$\pm$	0.02	&	19.33	$\pm$	0.03	&	i	\\
2456313.92	&	2013-01-21	&	18.58	$\pm$	0.02	&	19.29	$\pm$	0.03	&	i	\\
2456652.98	&	2013-12-26	&	18.45	$\pm$	0.01	&	19.28	$\pm$	0.02	&	i	\\
2456652.99	&	2013-12-26	&	18.53	$\pm$	0.01	&	19.30	$\pm$	0.02	&	i	\\
2456734.76	&	2014-03-18	&	18.64	$\pm$	0.02	&	\dots			&	i	\\
2456734.77	&	2014-03-18	&	18.59	$\pm$	0.02	&	\dots			&	i	\\
2456734.79	&	2014-03-18	&	18.67	$\pm$	0.02	&	\dots			&	i	\\
2455283.86	&	2010-03-28	&	15.38	$\pm$	0.01	&	\dots			&	z	\\
2455283.87	&	2010-03-28	&	15.37	$\pm$	0.01	&	\dots			&	z	\\
2455503.09	&	2010-11-02	&	18.30	$\pm$	0.02	&	18.84	$\pm$	0.03	&	z	\\
2455503.10	&	2010-11-02	&	18.33	$\pm$	0.02	&	18.89	$\pm$	0.03	&	z	\\
2455637.73	&	2011-03-17	&	19.86	$\pm$	0.09	&	19.26	$\pm$	0.04	&	z	\\
2455637.75	&	2011-03-17	&	19.53	$\pm$	0.07	&	19.34	$\pm$	0.05	&	z	\\
2455905.03	&	2011-12-09	&	16.52	$\pm$	0.01	&	\dots			&	z	\\
2455905.05	&	2011-12-09	&	16.59	$\pm$	0.01	&	\dots			&	z	\\
2456015.75	&	2012-03-29	&	17.31	$\pm$	0.01	&	18.67	$\pm$	0.03	&	z	\\
2456216.13	&	2012-10-15	&	17.81	$\pm$	0.02	&	\dots			&	z	\\
2456596.14	&	2013-10-30	&	16.80	$\pm$	0.01	&	18.68	$\pm$	0.02	&	z	\\
2456641.99	&	2013-12-15	&	17.31	$\pm$	0.01	&	18.76	$\pm$	0.03	&	z	\\
2455517.11	&	2010-11-16	&	17.42	$\pm$	0.02	&	18.57	$\pm$	0.05	&	y	\\
2455517.12	&	2010-11-16	&	\dots			&	18.66	$\pm$	0.05	&	y	\\
2455637.77	&	2011-03-17	&	18.91	$\pm$	0.09	&	18.88	$\pm$	0.08	&	y	\\
2455852.14	&	2011-10-17	&	\dots			&	18.21	$\pm$	0.04	&	y	\\
2456014.73	&	2012-03-28	&	16.77	$\pm$	0.02	&	18.54	$\pm$	0.06	&	y	\\
2456014.74	&	2012-03-28	&	16.74	$\pm$	0.02	&	18.50	$\pm$	0.06	&	y	\\
2456206.13	&	2012-10-05	&	16.96	$\pm$	0.02	&	18.54	$\pm$	0.06	&	y	\\
2456206.14	&	2012-10-05	&	17.03	$\pm$	0.02	&	18.53	$\pm$	0.06	&	y	\\
2456602.15	&	2013-11-05	&	16.23	$\pm$	0.01	&	\dots			&	y	\\
2456642.02	&	2013-12-15	&	16.84	$\pm$	0.02	&	18.32	$\pm$	0.05	&	y	\\
2456776.74	&	2014-04-29	&	17.03	$\pm$	0.01	&	18.52	$\pm$	0.04	&	y	\\
2456943.15	&	2014-10-12	&	17.23	$\pm$	0.03	&	18.62	$\pm$	0.07	&	y	\\
\enddata
\end{deluxetable}

\begin{deluxetable}{ccrr}
\tabletypesize{\scriptsize} 
\tablecaption{NEOWISE Photometry \label{tbl_wise_photometry}}
\tablewidth{0pt}
\tablehead{
\colhead{JD} & \colhead{Date [UT]} & \colhead{W1} & \colhead{W2} 
}
\startdata
 2455292.04 &  2010-04-05 &  8.610 $\pm$  0.015 &  7.513 $\pm$  0.015 \\
 2455483.53 &  2010-10-14 &  9.911 $\pm$  0.045 &  8.841 $\pm$  0.044 \\
 2456947.14 &  2014-10-16 & 11.019 $\pm$  0.018 &  9.809 $\pm$  0.017 \\
 2457115.87 &  2015-04-03 &  8.131 $\pm$  0.015 &  7.062 $\pm$  0.012 \\
 2457310.42 &  2015-10-14 & 10.449 $\pm$  0.017 &  9.395 $\pm$  0.015 \\
 2457474.94 &  2016-03-27 & 10.958 $\pm$  0.019 &  9.820 $\pm$  0.016 \\
 2457674.60 &  2016-10-13 &  9.460 $\pm$  0.016 &  8.270 $\pm$  0.013 \\
 2457835.48 &  2017-03-22 &  9.752 $\pm$  0.016 &  8.756 $\pm$  0.014 \\
 2458040.37 &  2017-10-13 & 10.324 $\pm$  0.017 &  9.173 $\pm$  0.015 \\
 2458196.11 &  2018-03-18 & 10.138 $\pm$  0.016 &  9.017 $\pm$  0.014 \\
 2458405.82 &  2018-10-14 &  9.314 $\pm$  0.016 &  8.321 $\pm$  0.013 \\
 2458563.12 &  2019-03-20 & 10.742 $\pm$  0.018 &  9.578 $\pm$  0.016 \\
 2458770.06 &  2019-10-13 & 10.541 $\pm$  0.017 &  9.274 $\pm$  0.015 \\
 2458927.25 &  2020-03-18 &  7.545 $\pm$  0.018 &  6.670 $\pm$  0.013 \\
 2459137.15 &  2020-10-14 &  9.668 $\pm$  0.017 &  8.709 $\pm$  0.014 \\
 2459292.54 &  2021-03-19 & 10.387 $\pm$  0.017 &  9.377 $\pm$  0.015 \\
 2459501.17 &  2021-10-13 &  7.928 $\pm$  0.016 &  6.884 $\pm$  0.012 \\
\enddata
\end{deluxetable}

\subsection{Near-infrared Spectroscopy}
\subsubsection{IGRINS}
NIR spectra of Gaia19fct were obtained with the Immersion GRating INfrared Spectrograph (IGRINS) installed on the 8.1\,m Gemini South telescope on 2020 November~19 and 23 (Program ID: GS-2020B-Q-218; PI: Park). IGRINS provides high-resolution (R=45,000) NIR spectra covering the full $H$ ($1.49-1.80$\,\um) and $K$ ($1.96-2.46$\,\um) bands with a single exposure \citep{yuk2010, park2014, mace2016}.
The spectrum was obtained with a slit scale of 0.34\arcsec~$\times$~5\arcsec. 
The signal-to-noise ratio (S/N) of $H$ and $K$ bands on November 19 are $\sim$190 and $\sim$270 and on November 23 are $\sim$137 and $\sim$201, respectively.
Gaia19fct was observed with several series of ABBA nodding observations at different positions on the slit to subtract the sky background. The exposure time of each nod observation was 300\,sec, and the total exposure time for 2020 November 19 and 23 were 3,600 and 2,400\,sec, respectively. Nearby A0 telluric standard stars (HIP~31226 and HIP~29933) were observed immediately after or before the observation of Gaia19fct for telluric correction.

The data reduction was done using the IGRINS pipeline \citep{lee2017} for flat-fielding, sky subtraction, correcting the distortion of the dispersion direction, wavelength calibration, and combining the spectra. 
Then, telluric correction was performed in the same manner as done in \citet{park2018}. 
For flux calibration, interpolated $H$ and $K$ band magnitudes between our RC80 $gVri$ bands and NEOWISE data observed on 2020 November 22 and 2020 October 14 were adopted since there is no recent NIR photometry close to the observing date.
The barycentric velocity ($V_{bary}$) was calculated by barycorrpy \citep{kanodia2018}, which is 20.22 and 19.12~km\,s$^{-1}$ for November 19 and 23, respectively.
The systemic velocity ($V_{LSR}$ = 13.25~km\,s$^{-1}$) obtained from APEX $^{13}$CO~$3-2$ data (Cruz-S\'aenz de Miera et al. submitted) is converted in the heliocentric system ($V_{helio}$ = 31.41~km\,s$^{-1}$) and used for the velocity correction. Finally, the $V_{bary}$ and systemic velocity ($V_{helio}$) correction was applied.

\subsubsection{NOTCam}
The intermediate resolution (R=2500) $H$ and $K$ bands spectra of Gaia19fct were acquired with NOTCam on the NOT on 2022 March 17 (Program ID: 65-111; PI: Cruz-S\'aenz de Miera). Gaia19fct was observed with an ABABAB pattern with exposure time of 210 and 294\,sec for $H$ and $K$ bands, respectively. The full pattern was observed twice for the $K$ band, thus the total exposure times for the $H$ and $K$ bands are 1,260 and 3,528\,sec, respectively. The nearby telluric standard star HD~44037 (B9~V) was observed right before the target observation for telluric correction. The spectroscopic observing log is listed in Table~\ref{tbl_info}.
The S/N of $H$ and $K$ band  spectra around 1.6 and 2.2\,\um\ are $\sim$24 and $\sim$20, respectively.

The raw data were reduced using {\sc IRAF} \citep{tody1986} for sky subtraction, flat-fielding, bad-pixel removal, aperture tracing, and wavelength calibration. Xenon lamp spectrum was used for the wavelength calibration. Hydrogen lines in the HD~44037 spectrum were removed, and then the spectrum was normalized. The target spectrum was divided by the normalized telluric spectrum to correct for the telluric lines. We used the $H$ and $K$ magnitudes observed with NOTCam on the same date for flux calibration. The $V_{bary}$ was calculated by barycorrpy \citep{kanodia2018} as $-23.07$\,km\,s$^{-1}$, and barycentric and systemic velocity correction ($V_{helio}$ = 31.41\,km\,s$^{-1}$) was applied.

\begin{deluxetable}{lcccrc}
\tabletypesize{\scriptsize} 
\tablecaption{Spectroscopic Observing Log \label{tbl_info}}
\tablewidth{0pt}

\tablehead{\colhead{Target} & \colhead{JD} & \colhead{Date} & \colhead{Band} & \colhead{Exp. time} & \colhead{Instrument} \\[-3mm]
& & \colhead{[UT]} & \colhead{} & \colhead{[sec]} &}
\startdata
Gaia19fct & 2459172.79 & 2020-11-19 & H, K & 300 $\times$ 12 & IGRINS \\
HIP~31226$^{a}$ & 2459172.84 & 2020-11-19 & H, K & 26 $\times$ 4 & IGRINS \\ 
Gaia19fct & 2459176.78 & 2020-11-23 & H, K & 300 $\times$ 8 & IGRINS \\
HIP~29933$^{a}$ & 2459176.76 & 2020-11-23 & H, K & 42 $\times$ 10 & IGRINS \\ 
Gaia19fct & 2459656.38 & 2022-03-17 & H & 210 $\times$ 6 & NOTCam \\
Gaia19fct & 2459656.43 & 2022-03-17 & K & 294 $\times$ 12 & NOTCam \\
HD~44037$^{b}$ & 2459656.36 & 2022-03-17 & H, K & 10 $\times$ 4 & NOTCam \\
\enddata
\tablenotetext{a}{Telluric standard star (A0 V) was observed right after or before the target to correct the telluric absorption features.}
\tablenotetext{b}{Telluric standard star (B9 V) was observed right before the target to correct the telluric absorption features.}
\end{deluxetable}

\section{Gaia19fct} \label{sec:gaia19fct}
\subsection{Location} \label{sec:distance}
Gaia19fct ($\alpha_{\rm J2000}$=07$^{\rm h}$\,09$^{\rm m}$\,21$\fs$39, $\delta_{\rm J2000}$=$-$10$^{\circ}$\,29$'$\,34$\farcs$55) is located close to the Galactic plane ($l$=224.30051, $b$=$-$0.84175) and lies towards the Canis Major OB1 (CMa OB1) association. According to \citet{sewilo2019}, Gaia19fct belongs to the ``CMa$-l$224" region (centered at ($l$, $b$) = (224.5, $-$0.65), see their Figures 3 and 4), where young protostars with outflows are found.
The median kinematic distance using the $^{12}$CO\,$1-0$ data is 0.92\,kpc, and about 99\% of targets in the CMa--$l$224 region are located between 0.5 and 1.3\,kpc distance \citep{sewilo2019}. 
This kinematic distance agrees with the Gaia DR2 distance of $1.14^{+1.01}_{-0.38}$\,kpc provided by \citet{bailer-jones2018} and $1.32 \pm 0.44$\,kpc calculated using the Gaia DR3 parallax \citep{gaiadr3}, within the uncertainties. 
Gaia DR3 provides the distance of Gaia19fct as $409.70^{+103.25}_{-86.50}$\,pc, however, the fractional parallax uncertainty\footnote{\url{https://gea.esac.esa.int/archive/documentation/GDR3/pdf/GaiaDR3_documentation_1.1.pdf}} is high (0.34). Therefore, we adopted the kinematic distance of 0.92\,kpc \citep{sewilo2019}, which is consistent with Gaia DR2 \citep{bailer-jones2018}, Gaia DR3 parallax distance, and previous studies about the distance of CMa OB1 association \citep{kaltcheva2000, gregorio-hetem2008} for the analysis in this work.

\subsection{Companion} \label{sec:companion}
\begin{figure}
    \centering
    \includegraphics[width=\columnwidth]{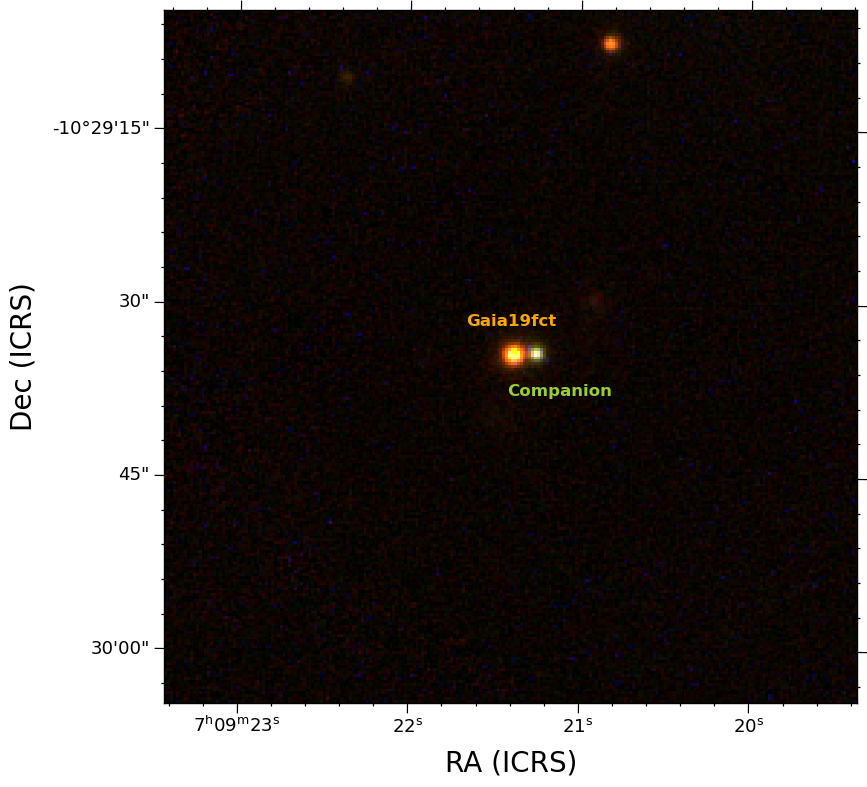}
    \caption{Pan-STARRS $giy$ color composite image of Gaia19fct. The image shows 1\arcmin$\times$1\arcmin{} area centered on Gaia19fct, and the companion is located 2\arcsec{} to the West.}
    \label{fig:color_composite}
\end{figure}

\autoref{fig:color_composite} shows the Pan-STARRS $giy$ color composite image of Gaia19fct (redder source) and its companion (bluer source). 
The companion is located $\sim$2\arcsec{} west of the target ($\sim$1840\,au in projection at a distance of 0.92\,kpc), and its coordinates are $\alpha_{\rm J2000}$=07$^{\rm h}$\,09$^{\rm m}$\,21$\fs$26, $\delta_{\rm J2000}$=$-$10$^{\circ}$\,29$'$\,34$\farcs$42 (Gaia DR3 source ID: 3046391410808013696).

We checked the distance of the companion to find out whether it is physically related to Gaia19fct. For this verification, we used the Gaia DR3 data, and the parallax distance of the companion ($1.38 \pm 1.29$\,kpc) is similar to Gaia19fct ($1.32 \pm 0.44$\,kpc), implying that Gaia19fct and the companion may form a physical pair.

We also examined the Pan-STARRS photometry of the companion separately from the target to check if it contaminates the photometry of Gaia19fct (\autoref{tbl_ps1_photometry}). The photometry result was constant within 0.3\,mag. In addition, the Gaia G-mag uncertainty vs. G-mag graph\footnote{\url{https://gea.esac.esa.int/archive/documentation/GEDR3/Data_processing/chap_cu5pho/cu5pho_sec_photProc/cu5pho_ssec_photVal.html}} shows that the companion is stable while Gaia19fct is variable. These results confirm that the companion is a stable source, not affecting our light curve analysis  for Gaia19fct.

\section{Results \& Analysis} \label{sec:results}

\subsection{Classification} \label{sec:classification}
Gaia19fct was classified as a Class~I YSO in the pre-outburst stage based on its spectral energy distribution (SED) and strong infrared excess \citep{miller2015}. Later, \citet{fischer2016} also suggested Gaia19fct as Class~I based on the infrared colors. \citet{sewilo2019} fitted the SED with a star and passive disk model and classified it as a disk-only source without an envelope. In order to revisit the evolutionary stage of Gaia19fct, we used several methods: bolometric temperature (\Tbol{}), spectral index ($\alpha$), and infrared colors.

\begin{figure}
    \centering
    \includegraphics[width=\columnwidth]{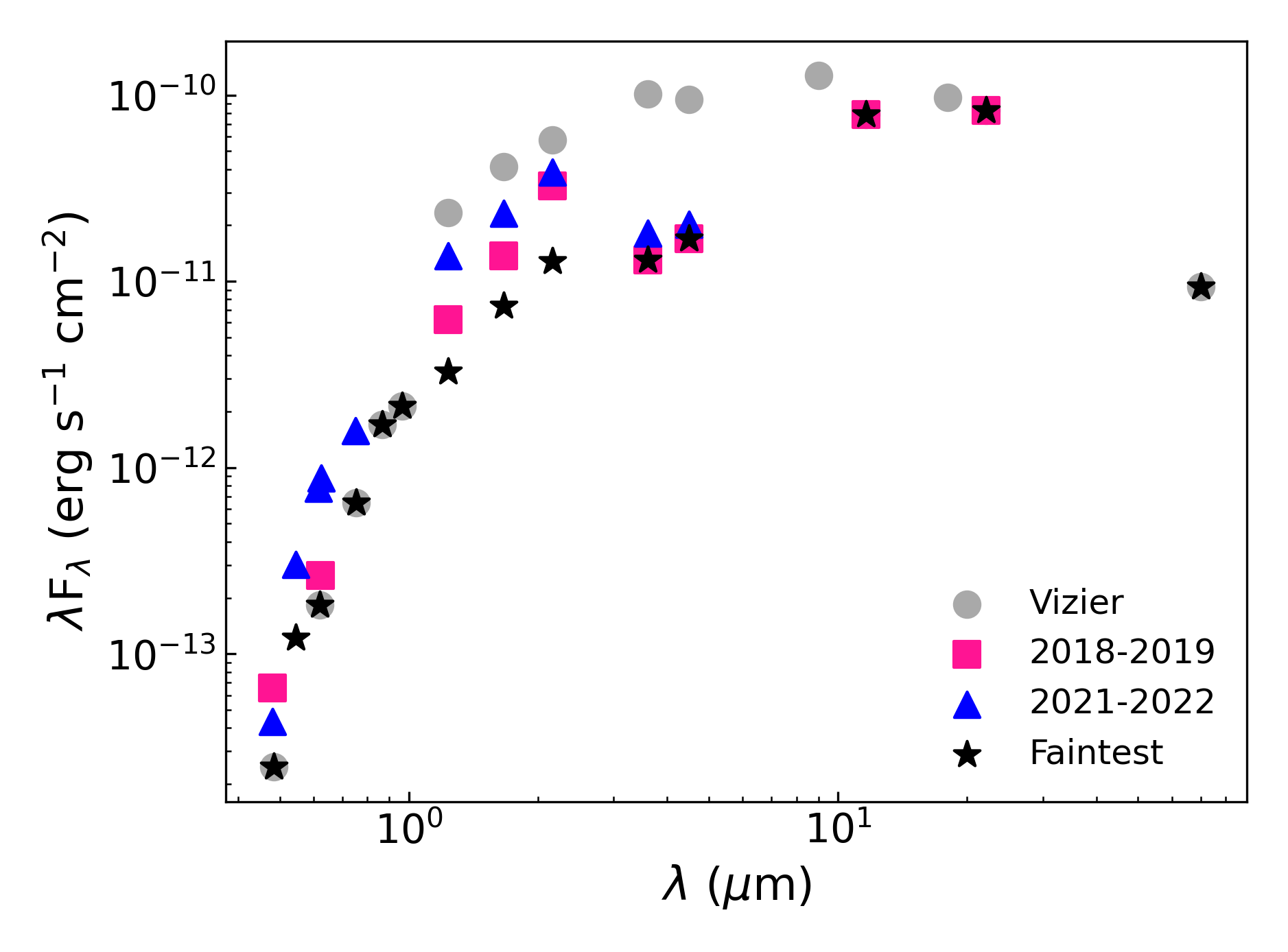}
    \caption{Spectral energy distribution of Gaia19fct. Different colors and symbols present different epochs. For the 2018-2019 period, we used the ZTF $gr$ and REM $JHK_{s}$ observed in 2018 November and August, respectively, and NEOWISE W1 and W2 observed in 2019 March. For the 2021-2022 period, we adopted the RC80 $gVri$, Gaia $G$, and REM $JHK_{s}$ observed in 2022 February and NEOWISE W1 and W2 obtained in 2021 March.
    \label{fig:sed}}
\end{figure}

We constructed the SED with several different datasets (\autoref{fig:sed}). First, we gathered the data observed before its discovery in 2015 (gray circles), and the data are from Pan-STARRS \citep[PS1;][]{chambers2016}, 2MASS \citep{cutri2003}, IRAC \citep{sewilo2019}, AllWISE \citep{cutri2014}, AKARI \citep{ishihara2010}, and PACS~70~\um\ \citep{sewilo2019}. The light curves (\autoref{fig:light_curve}) show at least five burst events, making a plausible assumption that this target had bursts before 2015. Therefore, we constructed additional SEDs based on the light curve.
Second, we used the ZTF $gr$, REM $JHK_{s}$, and NEOWISE W1 and W2 data observed in the 2018-2019 quiescent period (pink squares). 
Third, we collected the data observed in the 2021-2022 period with RC80 $gVri$, Gaia $G$, REM $JHK_{s}$, and WISE W1 and W2, the minimum phase between the two brightenings (blue triangles). 
Finally, we selected the faintest data (black stars) at each wavelength to assume the quiescent phase of Gaia19fct. 
The data observed at different epochs show different SED shapes, and we decided to use the 2018-2019 SED, which corresponds to the quiescent phase in the light curve, for the analysis.

We calculated ${L_{\rm bol}}$ and \Tbol{} for the 2018-2019 period, which is assumed to be the quiescent phase, by integrating the SED following the procedure described by \citet{chen1995}.
We used WISE W3 and W4 \citep{cutri2014}, and PACS~70\,\um\ \citep{sewilo2019} data for the longer wavelengths because only shorter wavelengths (up to 4.5\,\um) data were available.
The resulting ${L_{\rm bol}}$ and \Tbol{} are 5.6~$\pm$~1.6\,${L_{\odot}}$ and $481 \pm 10$\,K, respectively, when assuming a distance of 0.92\,kpc. The uncertainty of ${L_{\rm bol}}$ was estimated using a Markov Chain Monte Carlo approach. First, we created Gaussian distributions of each photometric measurement and the distance to the target, where each one of these distributions was made up of 50,000 elements. We used the values and uncertainties of each photometric point, and the distance, to inform the Gaussian mean and width. In the case where we could not obtain an uncertainty to a photometric measurement, we assumed uncertainty of 30\%. Second, we computed 50,000 values of ${L_{\rm bol}}$ using all the Gaussian distributions. Finally, we used this posterior distribution to calculate the 0.16, 0.50, and 0.84 quantiles and thus determine the value of ${L_{\rm bol}}$ and its uncertainties. Due to the small errors of the photometric points ($\sim$2\%), we infer that the uncertainty of ${L_{\rm bol}}$ is dominated by the distance and by the few photometric points with large uncertainties.
The uncertainty of \Tbol{} was calculated with photometric errors.
The calculated \Tbol{} is in the range of Class~I \citep[70\,K $\le$ \Tbol{} $\le$ 650\,K;][]{chen1995, evans2009}, suggesting that Gaia19fct is in the Class~I stage.
The $\alpha$ of pre-outburst SED between 2 to 24\,\um\ has been widely used to classify the evolutionary stage of YSOs \citep{lada1987, evans2009}. The $\alpha$ of Gaia19fct during the quiescent 2018-2019 period is about 0.4, which also corresponds to Class~I \citep[0.3 $\le$ $\alpha$;][and references therein]{evans2009}.

We also used 2MASS and WISE colors to check the YSO class of Gaia19fct. The MIR bands are related to a cooler circumstellar disk and envelope than the stellar photosphere, making MIR colors a useful indicator for YSO classification. 
Using these values of Gaia19fct, we found that it falls towards Class~I \citep{fischer2016, koening2014}. When both 2MASS and WISE colors are used, Gaia19fct falls between the Class~I and flat spectrum \citep{koening2014}. In addition, the location of Gaia19fct in the W1$-$W2 vs. W3$-$W4 plot is surrounded by Class~I and flat spectrum sources. 
The evolutionary stage of Gaia19fct varies between Class~I and flat spectrum depending on the methods. In this work, we will analyze Gaia19fct as a Class~I based on \Tbol{} and $\alpha$.

\subsection{Extinction} \label{sec:extinction}
We estimated an extinction (\Av) by comparing the spectrum of Gaia19fct to that of FU~Ori as done in \citet{connelley2018}. FU~Ori is known to have a low extinction \citep[\Av=1.7\,mag;][]{siwak2018, green2019, lykou2022}, making it a good comparison target. This method is valid for eruptive young stars whose optical and NIR spectra are dominated by an active accretion disk rather than the stellar photosphere \citep{hartmann1996, connelley2018, fischer2022}. The spectral continuum shape and lines observed in 2020 November are similar to those of FUors. Therefore, we assumed that the continuum shape of Gaia19fct observed in 2020 November is dominated by extinction, rather than the spectral type. The Gaia19fct spectrum was dereddened until it matched with that of FU Ori, and we matched the continuum shape of the spectrum, not the absolute flux level. As a result, we obtained \Av\ for 2020 November as $8 \pm 1$\,mag.

Extinction can also be estimated using the ratio of pairs of emission lines that share the same upper energy level \citep[][and references therein]{davis2011}.
Among several lines, only \hmol\ 1-0 S(1)/1-0 Q(3) lines were available in our 2020 November observation. 
The following equation is used to estimate the extinction \citep{davis2011}:
\begin{equation} \label{eq:av_davis2011}
    A_V = -114\times\log(0.704\times[I_{S1}/I_{Q3}]).
\end{equation}
The estimated \Av\ for 2020 November is $\sim$7.0\,mag, lower than that of \citet{sewilo2019} obtained by the SED fitting (\Av=10.2\,mag). 
In order to double-check the obtained \Av, we also used another equation from \citet{petersen1996}: 
\begin{equation} \label{eq:av_petersen1996}
    A_V = \frac{2.5\log(R_o/R_p)}{A_{\lambda_2}/A_V - A_{\lambda_1}/A_V}.
\end{equation}
R$_{o}$ is observed line ratio (I$_{Q3}$/I$_{S1}$) and R$_{p}$ is theoretical line ratio of $\sim$0.7 \citep{turner1977}. A$_{\lambda_1}$ and A$_{\lambda_2}$ are the wavelength of \hmol\ 1-0 Q(3) and \hmol\ 1-0 S(1), respectively.
To calculate A$_{\lambda_1}$/A${_V}$ and A$_{\lambda_2}$/A${_V}$, we used an extinction law of Equation~\ref{eq:extincion_law} from \citet{rieke1985}:
\begin{equation} \label{eq:extincion_law}
    A_\lambda=A_V(0.55\mu m/\lambda)^{1.6}.
\end{equation}
The calculated \Av\ using Equation~\ref{eq:av_petersen1996} is $\sim$7.2\,mag. 
The \Av\ calculated from the \hmol\ line ratio (mean value of the two equations is $7.1\pm0.2$\,mag) is lower than the target because \hmol\ lines originate from the jet. Still, this value agrees well with the uncertainty of \Av\ obtained with spectral comparison (\Av=$8 \pm 1$\,mag). Therefore, we assume that the \Av\ obtained by comparing with the FU~Ori spectrum is from Gaia19fct and use this value (\Av=$8 \pm 1$\,mag) for the analysis.

\subsection{Photometry}
\subsubsection{Light Curve} \label{sec:light_curve}
\begin{figure*}
    \centering
    \includegraphics[width=0.85\textwidth]{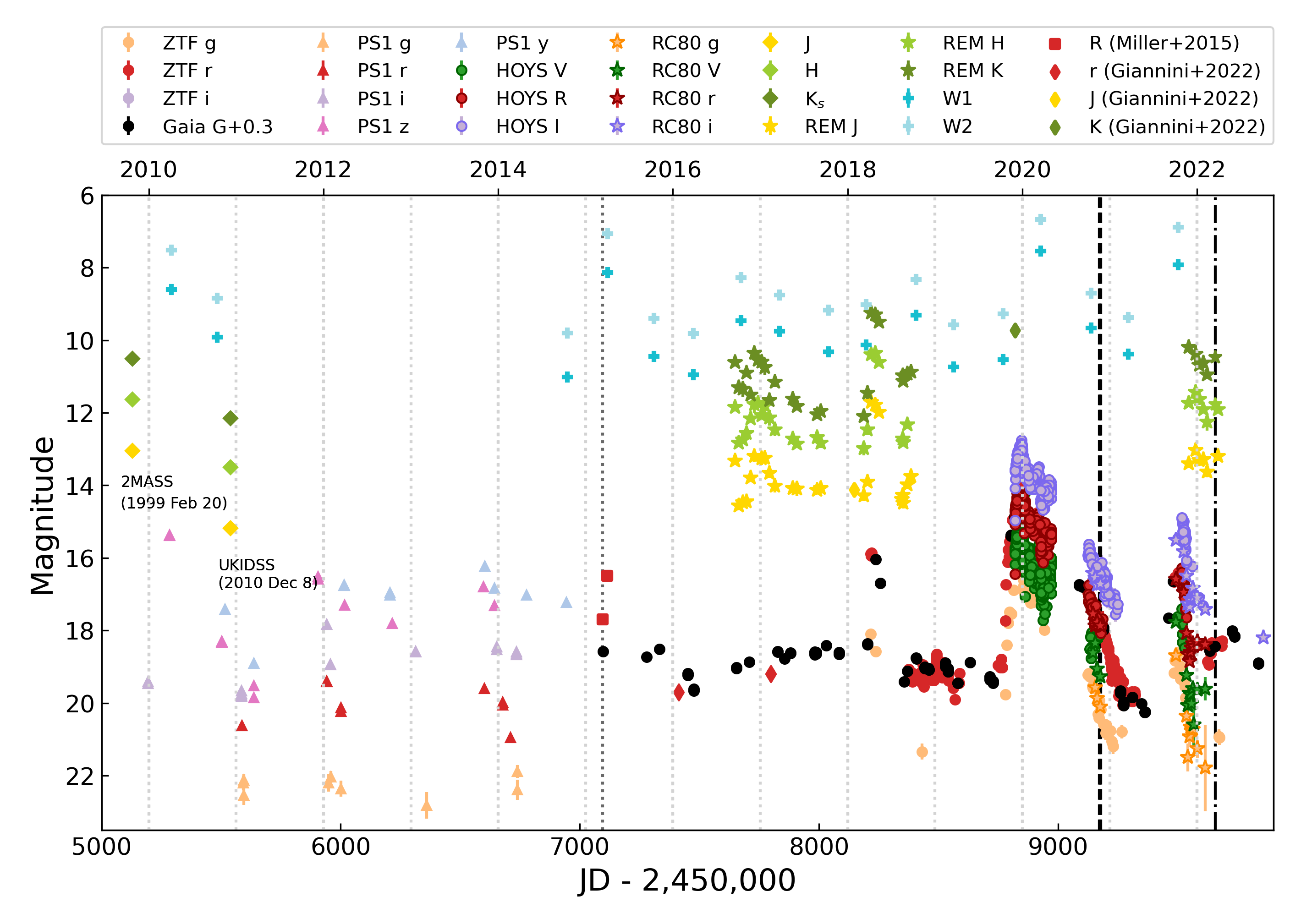}
    \includegraphics[width=0.85\textwidth]{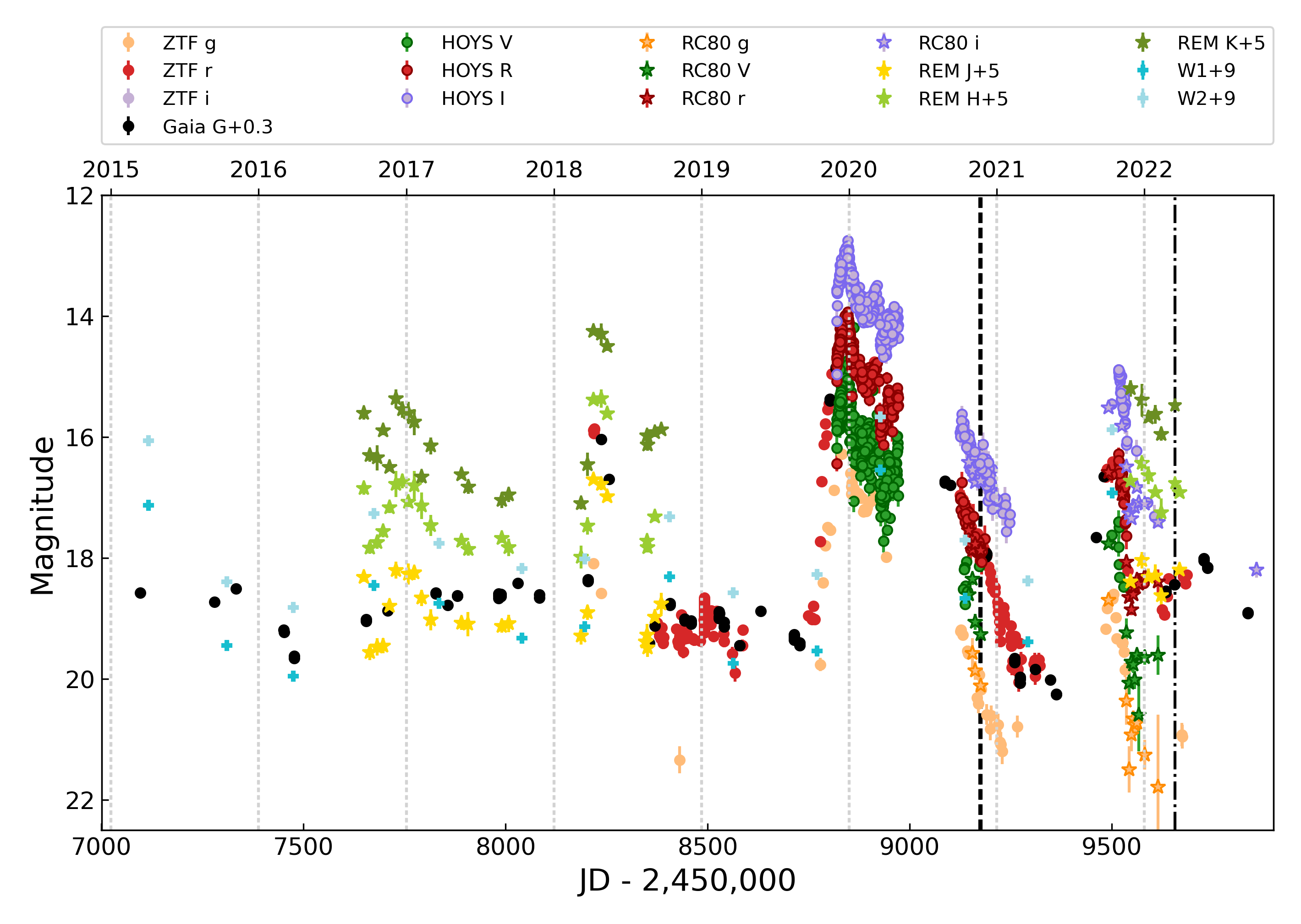}
    \caption{Light curve of Gaia19fct. ZTF \citep[$g$, $r$;][]{masci2018}, HOYS \citep[$V$, $R$, $I$;][]{froebrich2018, evitts2020}, WISE \citep[W1 and W2;][]{mainzer2011}, 2MASS \citep[$J$, $H$, $K_{s}$;][]{cutri2003}, UKIDSS \citep[$J$, $H$, $K_{s}$;][]{lawrence2007}, Pan-STARRS \citep[$g$, $r$, $i$, $z$, $y$;][]{chambers2016}, and Gaia archives ($G$), and our RC80 ($g$, $V$, $r$, $i$) and REM ($J$, $H$, $K_{s}$) monitoring observation data are used. Circle and star symbols present public domain data and our observations, respectively. Black dashed and dashed-dotted lines indicate IGRINS and NOTCam spectral observing dates, respectively. Uncertainties smaller than the symbol size are not presented.
    \label{fig:light_curve}}
\end{figure*}

\autoref{fig:light_curve} shows light curves of Gaia19fct.
The source has undergone brightening events at least four more times after its first discovery in 2015 \citep{miller2015}: 2016, 2018, 2019, and 2021. Again, the small brightening event in 2022 March recurred, and the recent data shows fading.
The amplitude and duration of the 2015, 2016, and 2018 events were moderate ($\le$~2.5\,mag) and short-lived, typical of EXors. 
The 2016 burst \citep[$\Delta V=\sim$1$-$2.5\,mag;][]{fischer2022} was only detected in our NIR observations with an amplitude of about one magnitude ($\Delta J\sim$1.2\,mag, $\Delta H\sim$0.8\,mag, and $\Delta K\sim$0.5\,mag), and the total duration of this event was only about two months.
Based on the ZTF $r$-band photometry, the fading rate of the 2018 event is about 0.021\,mag~day$^{-1}$ during about 170 days with $\Delta r$~$>$~3.5\,mag.
The amplitude of 2019 brightening is higher than 4.8\,mag, the largest burst in Gaia19fct, and is categorized as an outburst \citep[$\Delta V=\sim$2.5$-$6\,mag;][]{fischer2022}. The high amplitude is similar to what is typical for FUors \citep[][and references therein]{fischer2022} as well as the most powerful outbursts found in EXors (EX~Lup \citep{abraham2019, rigliaco2020}, Gaia20eae \citep{ghosh2022, csm2022}). The peak brightness was reached in 2019 December within $\sim$70\,days since the beginning of the outburst (2019 October), resulting in a steep rising rate of $-$0.072\,mag~day$^{-1}$. Then, the brightness faded again ($\Delta r$~$>$~5.3\,mag) with a rate of 0.015\,mag~day$^{-1}$ for about 357\,days. 
This target brightened again ($\Delta r$~$>$~3.5\,mag) in 2021 April with a rate of $-$0.019\,mag~day$^{-1}$ for about 191 days; the peak was in 2021 October ($r$=16.41\,mag), then faded with a rate of 0.044\,mag~day$^{-1}$ for about 49\,days ($\Delta r$~$>$~2.15\,mag).
Recently, Gaia19fct showed a small brightening variation ($\Delta G = 0.56$\,mag) in spring 2022 according to the Gaia $G$-band and ZTF $g$- and $r$-band photometry, and the latest Gaia $G$-band and RC80 $i$-band data show fading again ($\Delta G = 0.91$\,mag). The amplitude of the latest brightening event is less than the burst defined by \citet{fischer2022}.
The amplitude of the four bursts in 2015, 2016, 2018, and 2021 and one outburst event in 2019 resembles EXors and FUors, respectively. The duration of each event lasts less than a year, which is more similar to EXors.
Overall, the moderate amplitudes and short-lived time scales of the bursts of Gaia19fct make it more similar to EXors \citep{fedele2007, fischer2022}.

The WISE light curves show a brightness change trend with about 1800\,days cycle between 2010 and 2021. 
The detailed analysis is not possible because of the coarse data coverage, but the brightness peaks periods in 2015, 2016, 2020, and 2021 match the optical and NIR light curves. 
While no contemporary optical or NIR data are available from 2010, the difference between the two WISE observations taken about half a year apart strongly suggests that there was a brightening event also in 2010.

\subsubsection{Search for Small-scale Quasi-periodic Light Changes} \label{sec:small_scale}
To get an insight into the small-scale variability, we performed a frequency analysis of the most numerous HOYS $RI$ and ZTF $r$-band data obtained during the maxima and the fading stages. We aimed to find small-scale periodic or quasi-periodic light variations other than the large-scale ones leading to the major outbursts. Detection of time-coherent small-scale variability can be used to constrain the inner disk dynamic or at least the outer environment in FUors and EXors. The best established results are usually obtained when the light curves are continuously gathered by a spacecraft \citep{siwak2013, siwak2018, siwak2020, hodapp2019, szabo2021}, but this kind of analysis is also possible for well-sampled ground-based light curves \citep{green2013b, hackstein2015, baek2015, ghosh2022}.

As a result, we obtained that except for a few random-like variations directly visible in the light curves, there is no significant variability that would last for at least a few consecutive cycles. This is reminiscent of the situation in V1057~Cyg \citep{szabo2021} and V1515~Cyg \citep{szabo2022}, where the inner disk light is strongly reprocessed by the obscuring surrounding envelope \citep{szabo2021}.

\subsubsection{Color Variation}
\begin{figure*}
    \centering
    \includegraphics[width=0.45\textwidth]{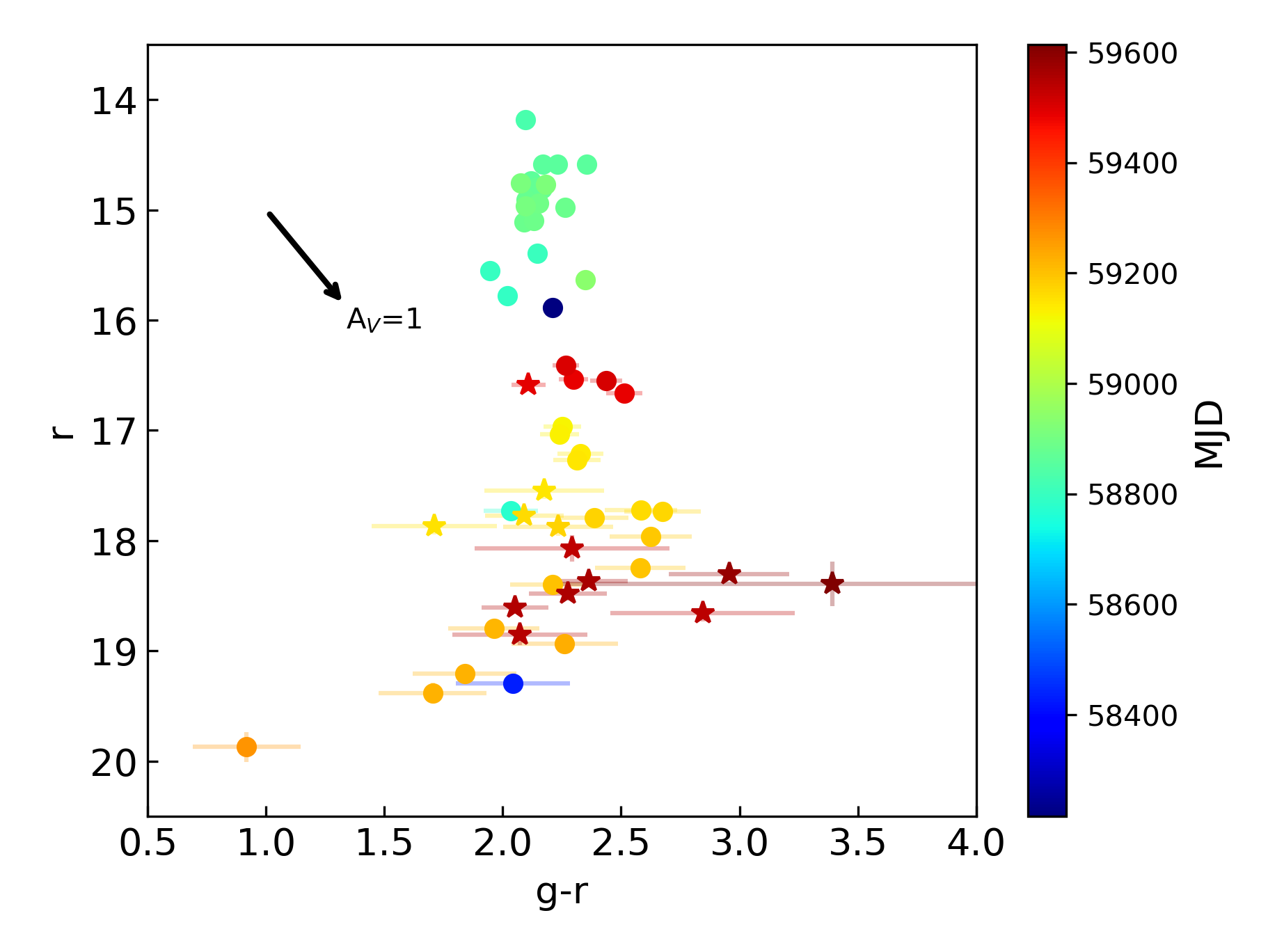}
    \includegraphics[width=0.45\textwidth]{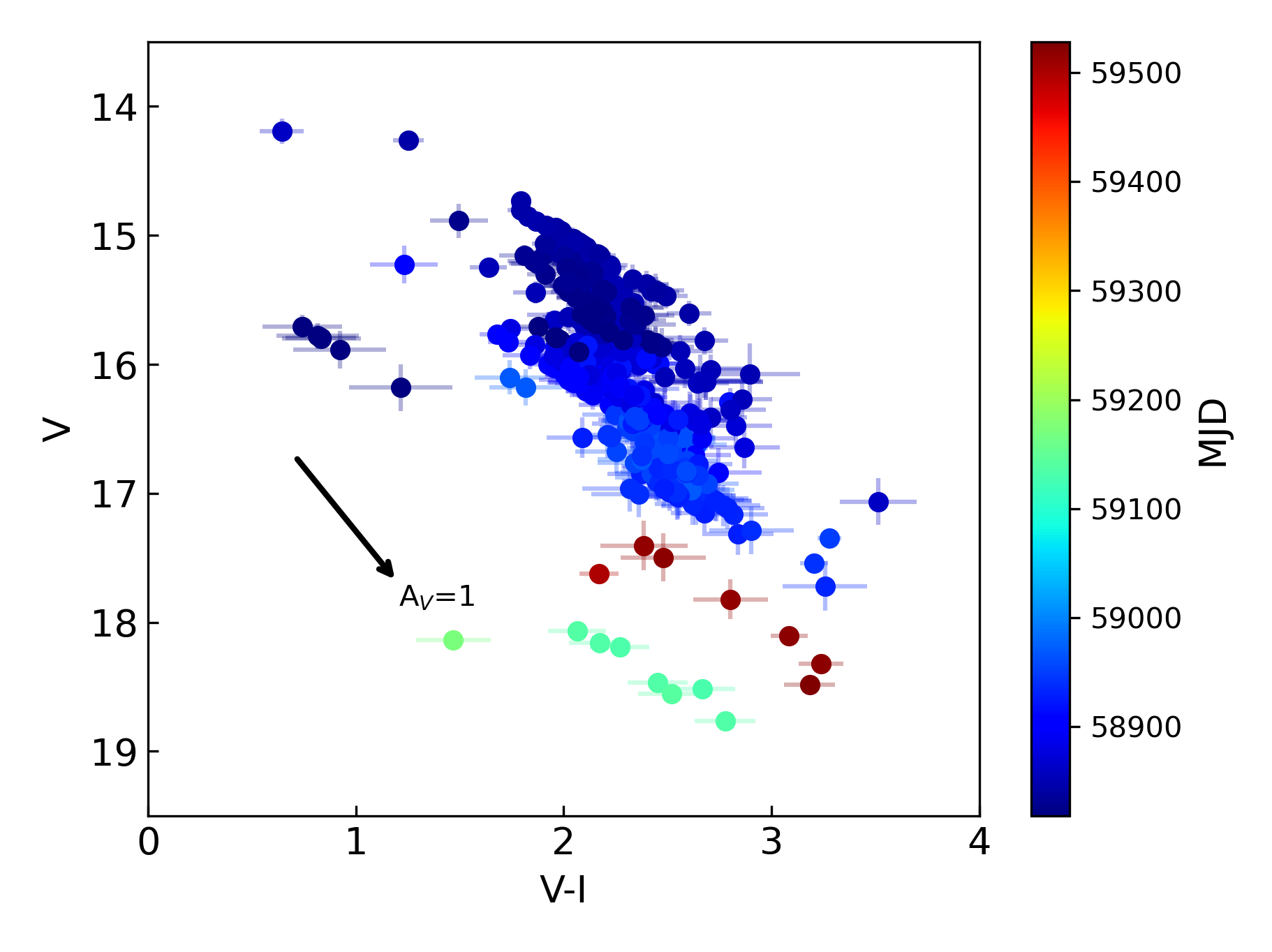}
    \includegraphics[width=0.45\textwidth]{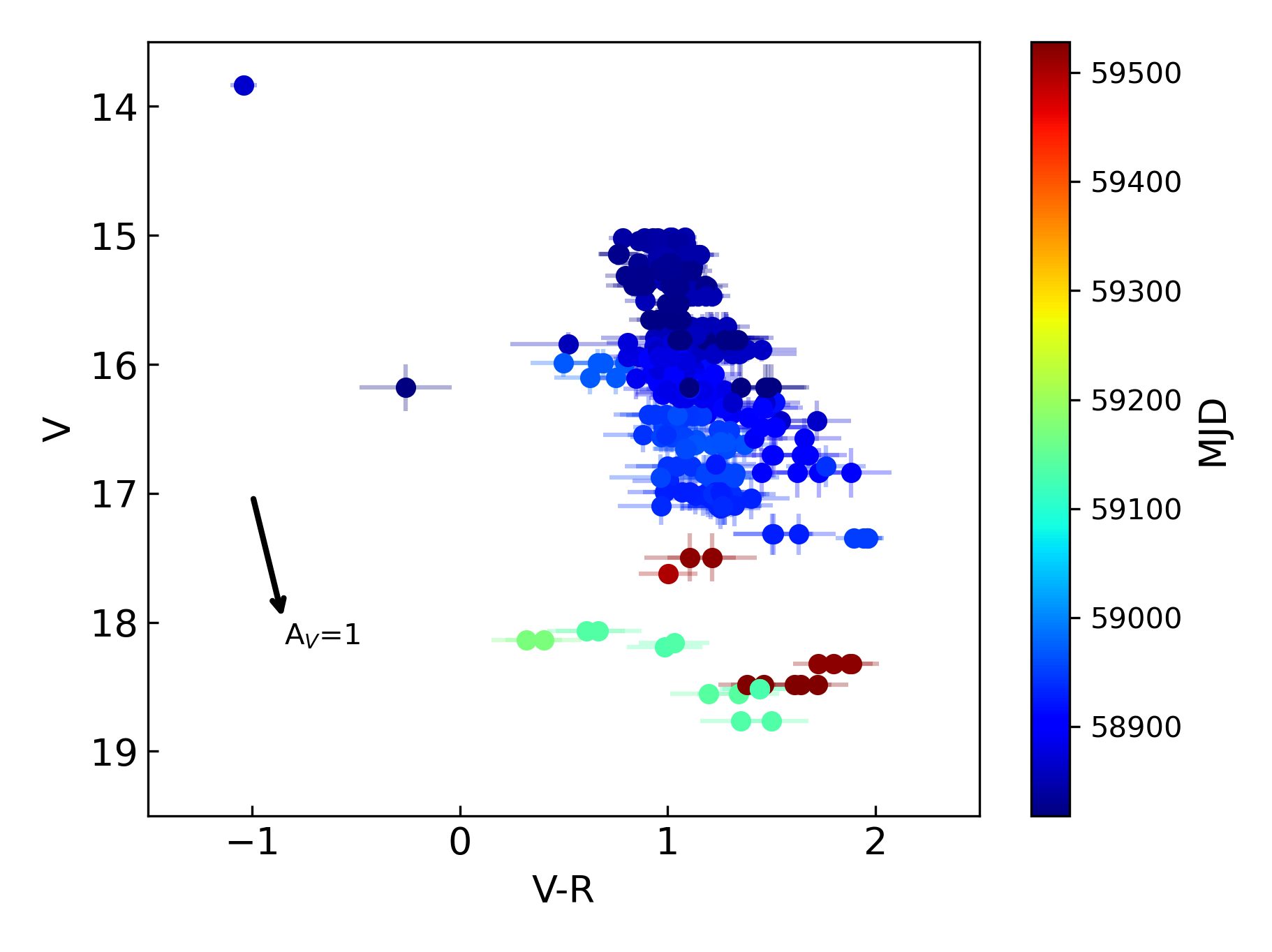}
    \includegraphics[width=0.45\textwidth]{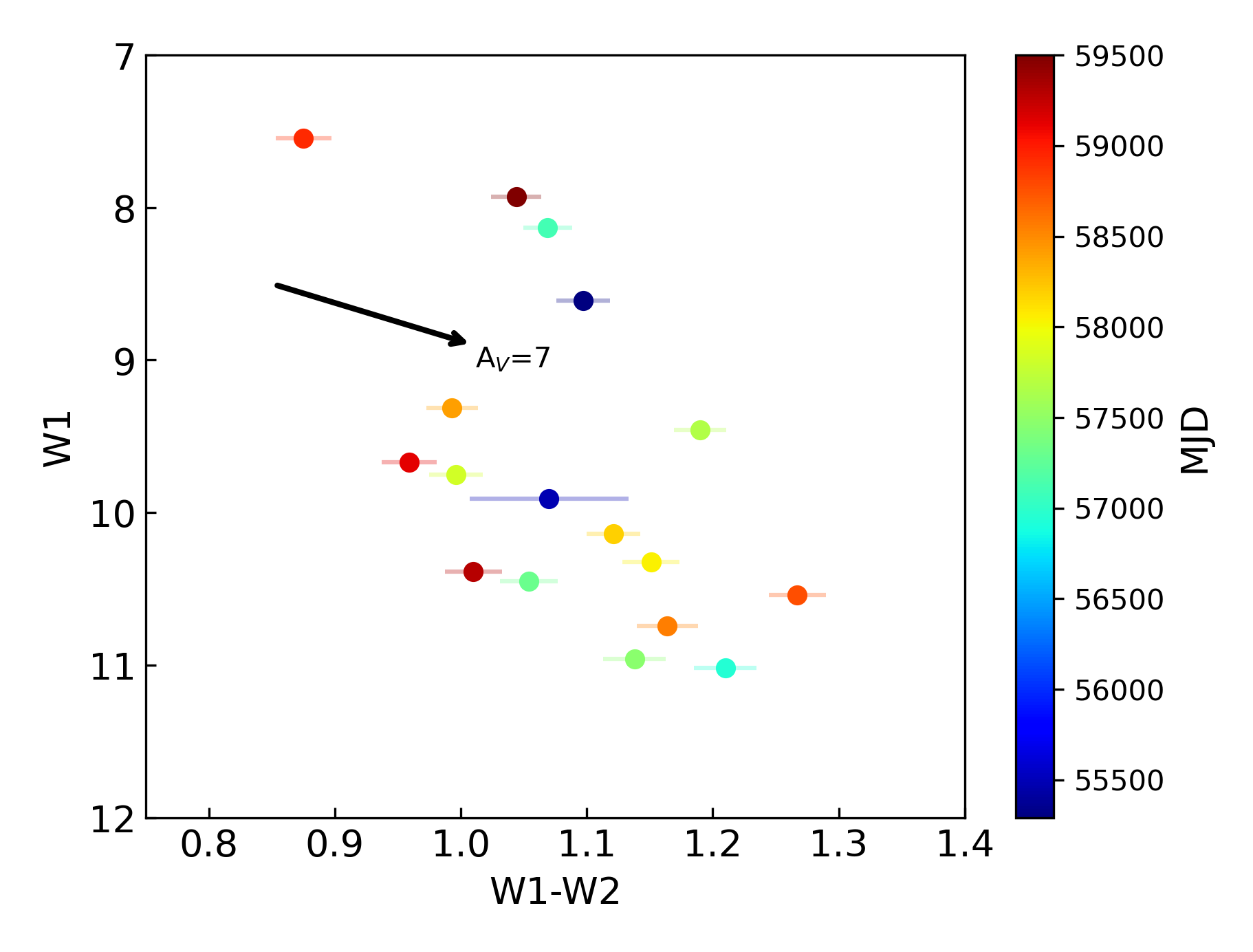}
    \includegraphics[width=0.45\textwidth]{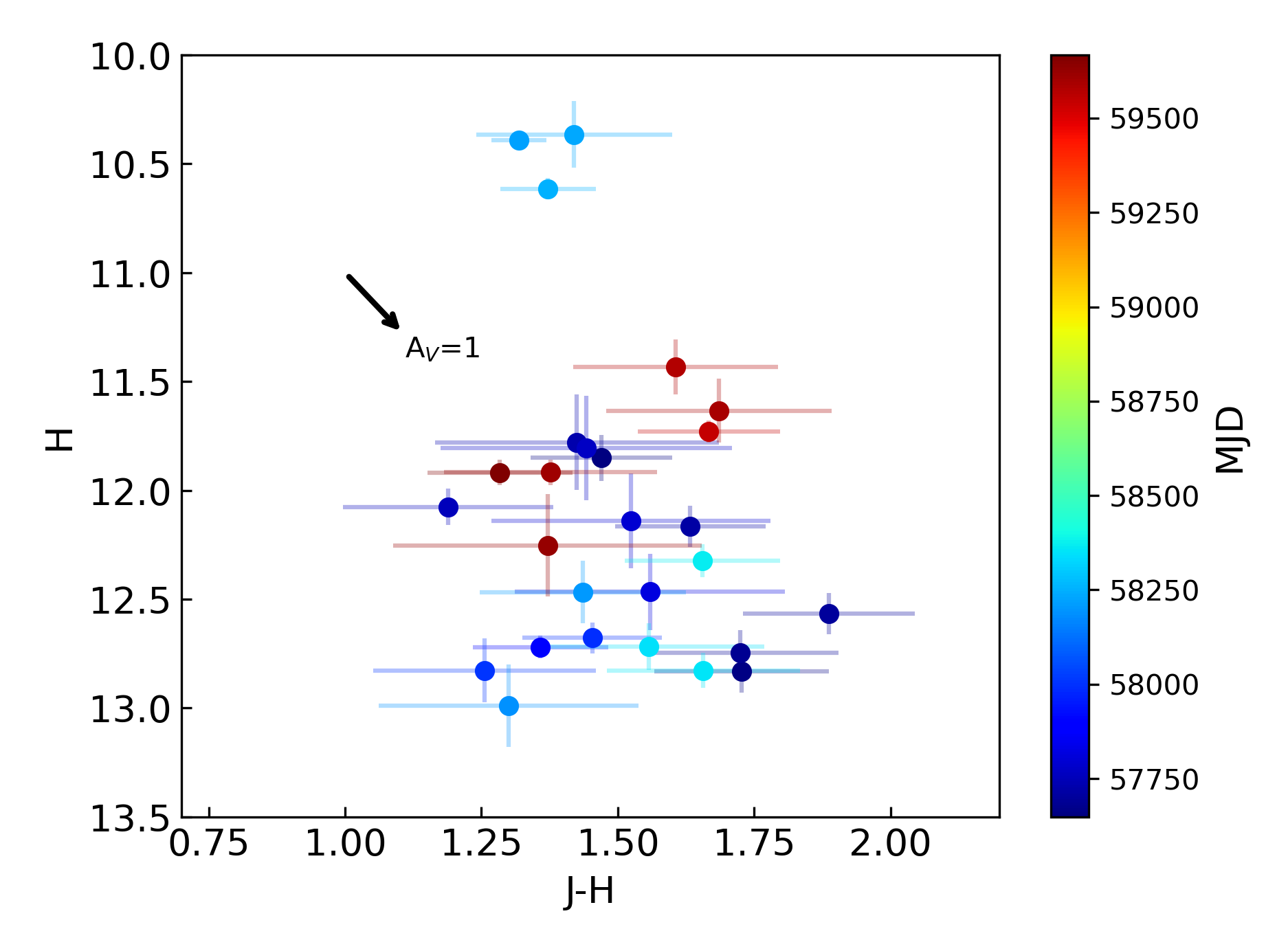}
    \includegraphics[width=0.45\textwidth]{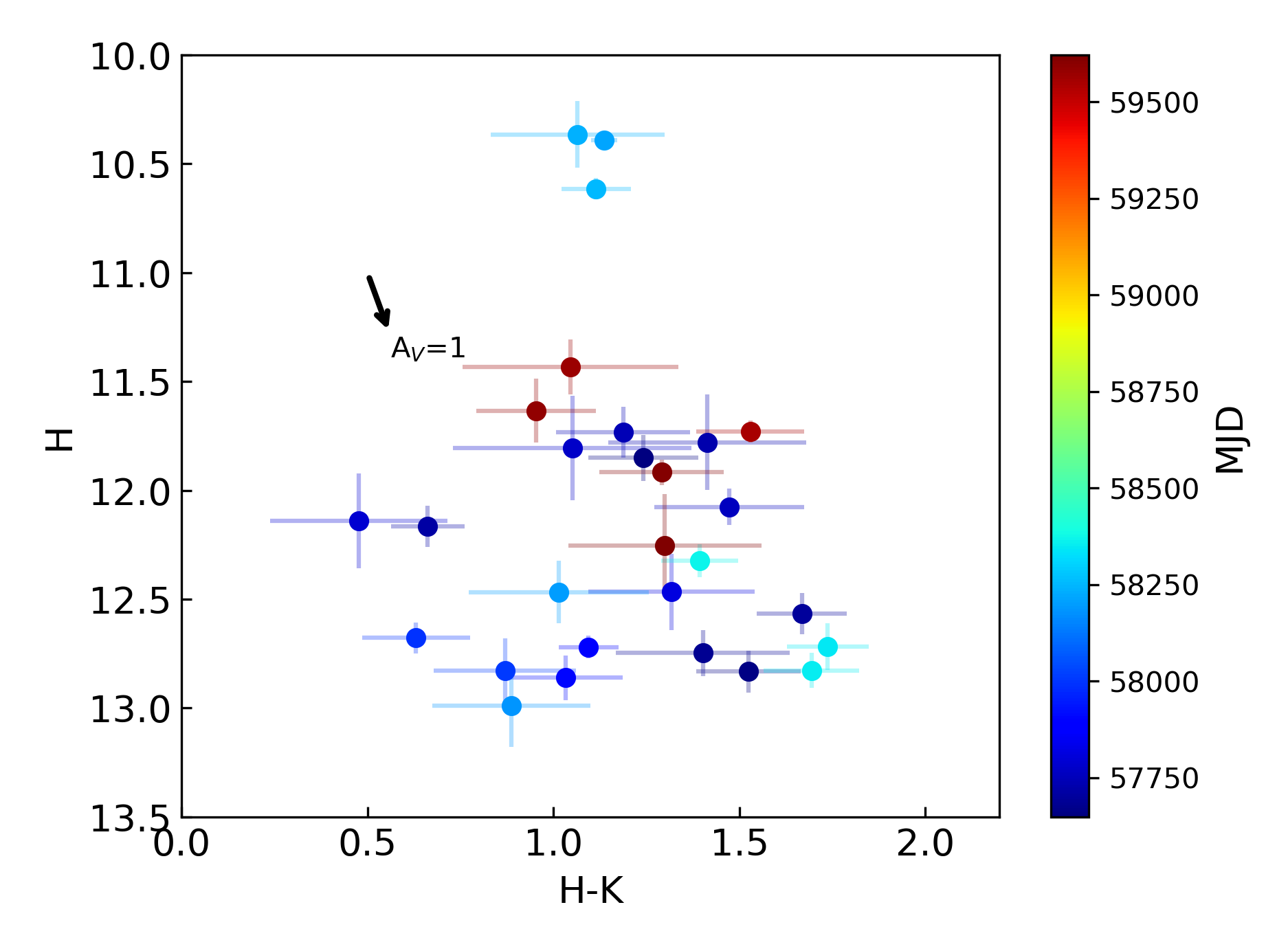}
    \caption{
    Color-magnitude diagrams of Gaia19fct. Upper left: $r$ vs. $g-r$ diagram based on the ZTF (circle) and RC80 (star) data. Upper right and middle left: $V$ vs. $V-I$ and $V$ vs. $V-R$ diagrams based on HOYS data. Middle right: W1 vs. W1$-$W2 diagram based on NEOWISE data. Lower left and right: REM H vs. $J-H$ and $H$ vs. $H-K$ diagrams.
    \label{fig:cmd}}
\end{figure*}

\autoref{fig:cmd} shows color-magnitude diagrams (CMDs) of Gaia19fct using different photometric bandpasses. 
The upper left panel presents $r$ vs. $g-r$ CMD constructed using ZTF and our RC80 data and shows gray dimming since late 2019. 
The upper right and middle left panels show HOYS $V$ vs. $V-I$ and $V$ vs. $V-R$ CMDs. In both CMDs, all small color variations seem to follow the extinction path. However, the overall trend from 2019 to 2022 shows gray evolution, which is also seen in the NEOWISE and $JHK_{s}$ color variations (middle right and lower panels). The gray variability in the optical and IR suggests that mechanisms other than extinction change cause the color variation of the 2019 and 2021 bursts.

\subsection{Near-infrared spectroscopy}

\begin{figure*}
    \centering
    \includegraphics[width=\textwidth]{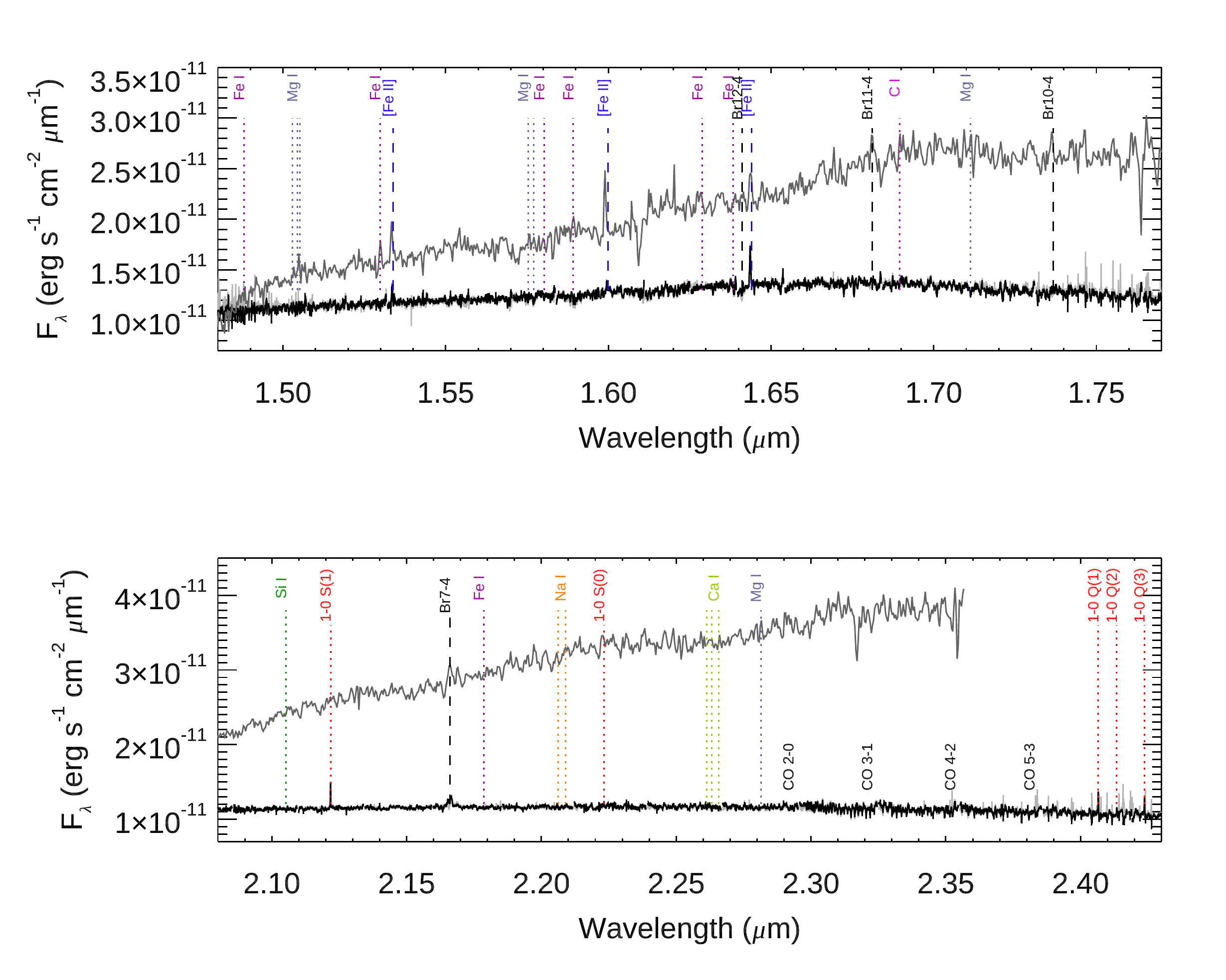}
    \caption{H and K band spectra of Gaia19fct observed on 2020 November 19 (black), 23 (light gray), and 2022 March 17 (dark gray). Telluric correction of the second observation (light gray) was not perfect, therefore, emission profiles remained at the edge of each band, especially longer than 2.38\,\um. Different colors indicate different spectral lines.
    \label{fig:1d_spec}}
\end{figure*}

\autoref{fig:1d_spec} shows our NIR spectra obtained in 2020 November 19 (black), 23 (light gray) and 2022 March 17 (dark gray). 
In 2020 November, $H$ band continuum shape of Gaia19fct is triangular, and $K$ band continuum shape is almost flat and decreases from 2.3\,\um. The $H$ and $K$ band continua shapes of Gaia19fct are more similar to FUors than FUor-like or Peculiar objects \citep{connelley2018}. While in 2022 March, both $H$ and $K$ band continua increase toward longer wavelengths, and the steeply rising $K$ band continuum shape is typical of Class~I sources \citep{beck2007, connelley2010}.

Most FUors show absorption dominated spectra formed by a viscously heated inner disk midplane \citep[][and references therein]{connelley2018}, and only a few FUors show emission lines, see for example, V2494~Cyg \citep{magakian2013}, V960~Mon \citep{takagi2018, park2020}, V346~Nor \citep{kospal2020}, and V1057~Cyg \citep{szabo2021}. On the other hand, EXors show emission dominated spectrum formed by a magnetospheric funnel \citep[][and references therein]{fischer2022}.

Our NIR spectra show both absorption and emission lines. In our first observation epoch with IGRINS in 2020 November, atomic metal lines are observed in absorption, characteristics of FUors. Several emission lines are also observed, including [\FeII], \hmol, and Br series, and these lines are associated with jets/outflows and accretion. 
CO overtone bandheads are observed weakly in emission, and the overall CO features show a superposition of emission and absorption.
These emission lines are typically observed in EXors or embedded YSOs. The observed absorption and emission lines do not vary in the two observations within four days except in the Br series. Therefore, the two IGRINS spectra were combined by observation error weighting to increase the S/N, and the combined spectrum was used for the analysis. The second epoch observation with NOTCam in 2022 March is dominated by emission lines, including [\FeII] and Br series. In this observation, \hmol\ and CO features are not observed: may be absent or too weak to be detected. The continuum shapes and spectral lines of the two epochs varied depending on the observation dates. The overall spectral characteristics show similarities with both FUors and EXors.

\subsubsection{Equivalent width} \label{sec:ew}
We measured the equivalent width (EW) of relatively strong and isolated lines to study how the spectral lines changed. Each line was fitted with Gaussian to define the integration range, and then 3$\sigma$ criterion was used for the integration. EW was estimated with a Monte Carlo method with random Gaussian errors multiplied by the observation errors. EW of each line was measured 100 times, and the standard deviation of all measurements was adopted as the uncertainty of the EW.
In the case of the Br series observed on 2020 November 19 and 23, EW was measured for both spectra since these Br series only varied between two IGRINS observations. We used the weighted mean spectrum of 2020 November for the other lines, which did not vary between the two observations. Measured EWs are listed in Table~\ref{tbl_ew}. 

\begin{deluxetable}{lccc}
\tabletypesize{\scriptsize} 
\tablecaption{Equivalent Widths \label{tbl_ew}}
\tablewidth{0pt}
\tablehead{\colhead{Transition} & \colhead{Wavelength} & \colhead{EW (2020 November)} & \colhead{EW (2022 March)} \\ [-2mm]
\colhead{} & \colhead{(\um{})} & \colhead{(\AA{})} & \colhead{(\AA{})}
}
\startdata
Br 12-4 &  1.641 &                                         \dots &   -0.876 $\pm$ 0.177 \\
Br 11-4 &  1.681 &      -0.301 $\pm$ 0.008$^{a}$     0.325 $\pm$ 0.009$^{b}$ &   -1.832 $\pm$ 0.242 \\
Br 10-4 &  1.737 &                                         \dots &   -2.294 $\pm$ 0.319 \\
Br 7-4  &  2.166 &     -2.252 $\pm$ 0.010$^{a}$     -1.633 $\pm$ 0.012$^{b}$ &   -1.434 $\pm$ 0.363 \\
CO 2-0  &  2.293 &                            -0.849 $\pm$ 0.022 &                \dots \\
1-0 S3  &  1.958 &                            -1.025 $\pm$ 0.028 &                \dots \\
1-0 S2  &  2.034 &                            -0.176 $\pm$ 0.002 &                \dots \\
1-0 S1  &  2.122 &                            -0.534 $\pm$ 0.002 &                \dots \\
1-0 S0  &  2.223 &                            -0.158 $\pm$ 0.002 &                \dots \\
1-0 Q1  &  2.407 &                            -0.510 $\pm$ 0.002 &                \dots \\
1-0 Q2  &  2.413 &                            -0.133 $\pm$ 0.002 &                \dots \\
1-0 Q3  &  2.424 &                            -0.451 $\pm$ 0.003 &                \dots \\ \relax
[Fe\,{\scriptsize II}]  &  1.534 &                            -0.460 $\pm$ 0.005 &   -2.350 $\pm$ 0.297 \\ \relax
[Fe\,{\scriptsize II}]  &  1.600 &                            -0.334 $\pm$ 0.003 &   -2.074 $\pm$ 0.329 \\ \relax
[Fe\,{\scriptsize II}]  &  1.644 &                            -1.105 $\pm$ 0.003 &   -0.684 $\pm$ 0.185 \\
\enddata
\tablenotetext{a}{Observed on 2020 November 19}
\tablenotetext{b}{Observed on 2022 November 23}
\end{deluxetable}

\subsubsection{Brackett series}
\begin{figure*}
    \centering
    \includegraphics[width=\textwidth]{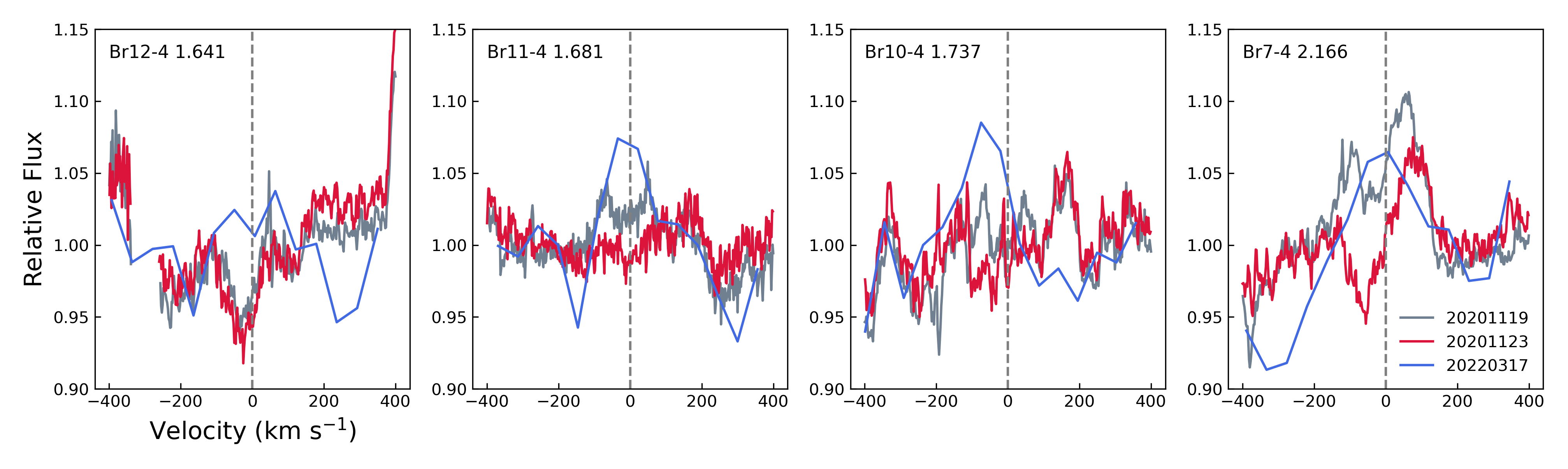}
    \caption{Observed Brackett series of Gaia19fct. Gray, red, and blue lines represent the data observed on 2020 November 19, 23, and 2022 March 17.
    \label{fig:Br_series}}
\end{figure*}

Four Br series lines are observed in emission (\autoref{fig:Br_series}), where the {H\,\footnotesize{I}} emission lines are typically found in EXors \citep{ lorenzetti2009}.
The Br lines are the only ones that show variation between the two IGRINS observations, therefore, we measured the EWs of each line (Section~\ref{sec:ew} and Table~\ref{tbl_ew}). The strengths of these lines on 2020 November 23 are weaker than in the spectrum taken four days earlier. Especially, the blue-shifted absorption component of Br$\gamma$ became stronger, indicating that the strength of wind increased within four days. The Br\,12-4 and Br\,10-4 lines are marginally detected in the 2020 November observations but observed firmly in 2022 March. The EW of the Br\,11-4 increased in 2022, while the EW of the Br$\gamma$ decreased in 2022. The EW of observed spectral lines varies depending on the observation epoch. These spectral line variations are consistent with previous studies \citep{hillenbrand2019, giannini2022}, which show different line profiles depending on the brightness of Gaia19fct.

\subsubsection{\hmol\ lines} \label{sec:h2}
\begin{figure*}
    \centering
    \includegraphics[width=\textwidth]{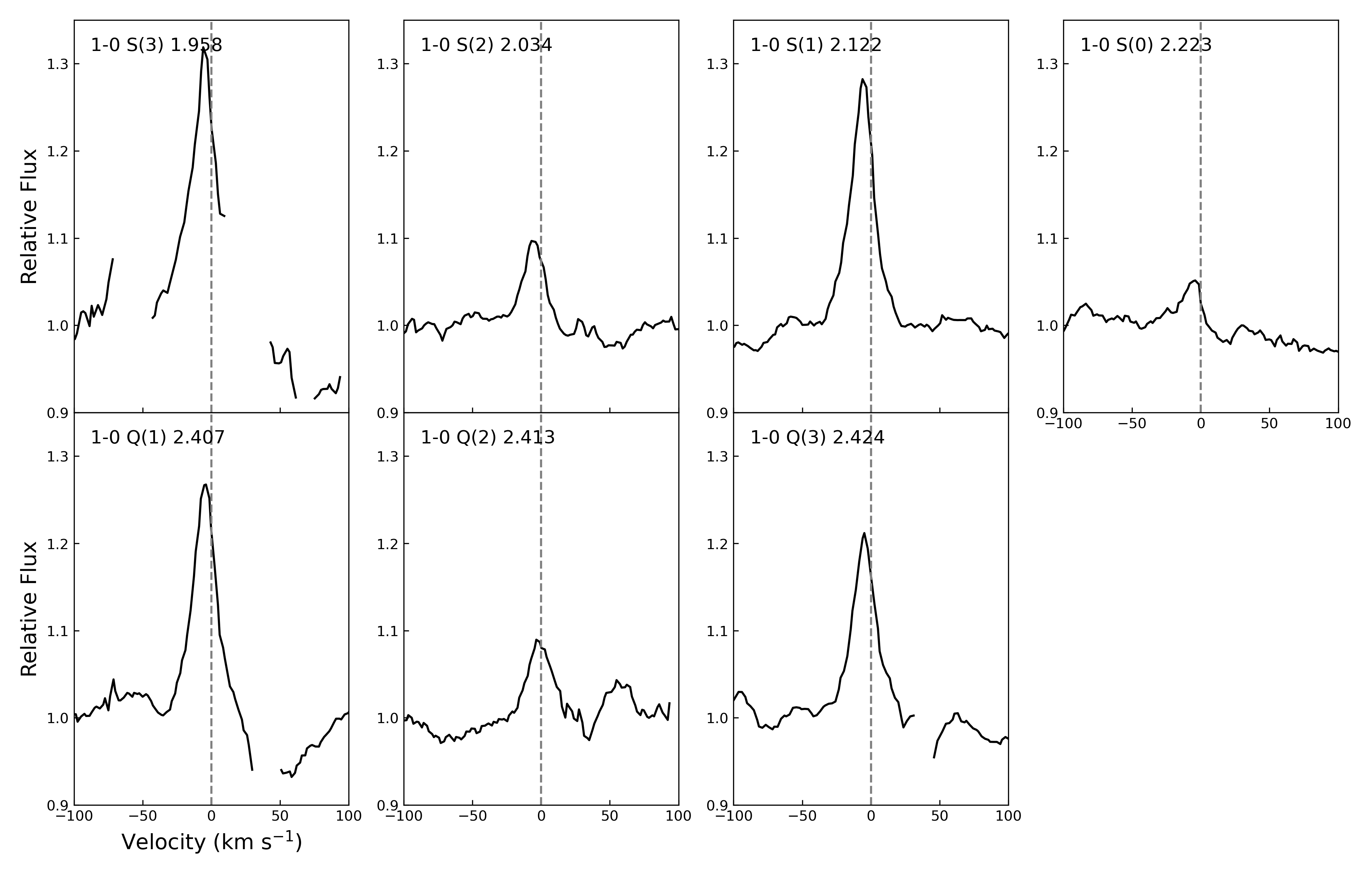}
    \caption{Observed \hmol\ lines in 2020 November. The part where the telluric correction is not perfect is missing, e.g., 1-0 S(3) 1.958\,\um.
    \label{fig:h2_lines}}
\end{figure*}

Several \hmol\ emission lines were observed in 2020 November (\autoref{fig:h2_lines}), which are rarely detected in FUors or EXors. Only about 10\% of eruptive stars showed \hmol\ rovibrational transitions: V1647~Ori \citep{aspin2011}, PTF~10nvg \citep{hillenbrand2013}, Gaia19ajj \citep{hillenbrand2019_gaia19ajj}, V960~Mon \citep{park2020}, Gaia19bey \citep{hodapp2020}, V346~Nor \citep{kospal2020}, and V899~Mon \citep{park2021_v899mon}.
The mean and standard deviation of the peak velocity is $-5\pm1$\,km\,s$^{-1}$, slightly blue-shifted with respect to the systemic velocity. This blue-shifted velocity indicates that these lines are formed by an outflowing wind \citep{vandenAncker1999, fernandes2000, nisini2002, davis2003, davis2010, greene2010, davis2011, bally2007, bally2016}. 

\begin{figure}
    \centering
    \includegraphics[width=\columnwidth]{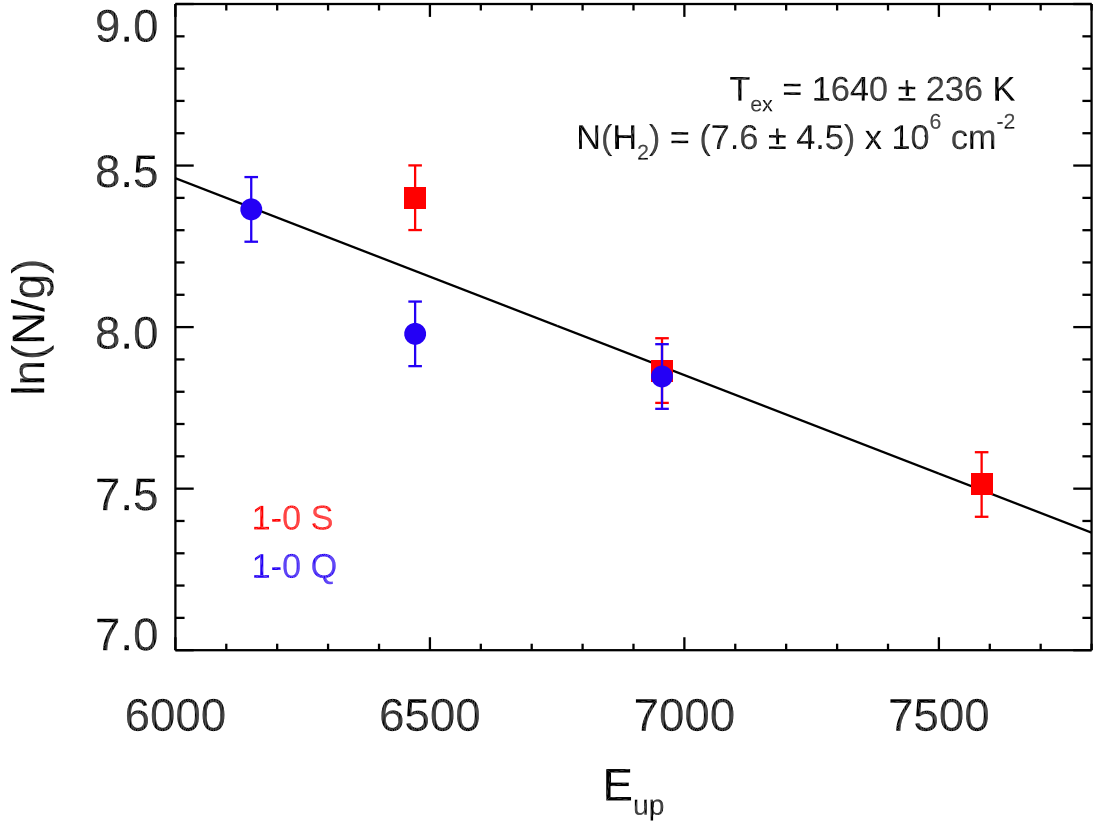}
    \caption{Excitation diagram of \hmol.}
    \label{fig:excitation_diagram}
\end{figure}

We constructed an excitation diagram (\autoref{fig:excitation_diagram}) to estimate the gas excitation temperature (\Tex) and the column density ($N_{\rm H2}$) of \hmol\ lines \citep{vandenAncker1999, fernandes2000, nisini2002, davis2011, oh2018, park2021_v899mon}.
The excitation diagram can be fitted by a single straight line if the gas is thermalized with a single temperature. \Tex\ can be obtained by the reciprocal slope of the fitted line, and the y-intercept can be used to calculate the $N_{\rm H2}$.
For this analysis, we adopted the \Av=$8 \pm 1$\,mag, and the line flux and line flux uncertainty were measured using the same method as EW. 
The estimated uncertainty of the line flux is about 1\%, which might be underestimated. We considered the possible uncertainties of \Av, continuum fitting, and telluric correction and then assumed a 10\% uncertainty for the line flux.
In addition, we used only six lines, excluding the 1-0~S(3)\,1.958\,\um\ line located in the crowded telluric region. 
The obtained \Tex\ and $N_{\rm H2}$ are 1640~$\pm$~236\,K and (7.6~$\pm$~4.5)~$\times$~10$^6$\,cm$^{-2}$, respectively. The obtained \Tex\ is lower than found for other eruptive young stars \citep[V346~Nor and V899~Mon;][]{kospal2020, park2021_v899mon}, suggesting that the \Tex\ of shock-heated region in Gaia19fct is lower than in these two targets.

\subsubsection{[\FeII] lines}
\begin{figure*}
    \centering
    \includegraphics[width=\textwidth]{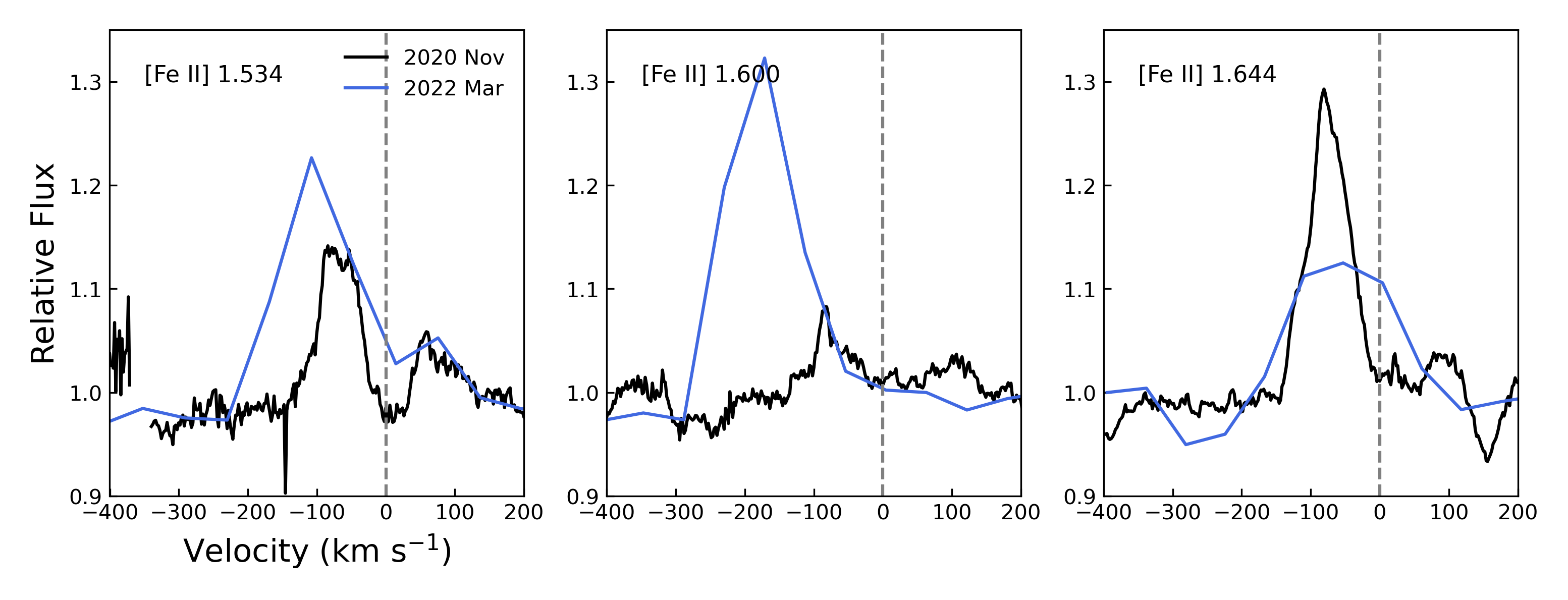}
    \caption{[\FeII] emission lines observed in Gaia19fct. Black and blue lines present the spectrum observed in 2020 November and 2022 March, respectively.
    \label{fig:feii_lines}}
\end{figure*}

\autoref{fig:feii_lines} shows the comparison of  forbidden [\FeII] lines observed in 2020 November (black) and 2022 March (blue). The [\FeII] lines in the second epoch became stronger and faster than in the first epoch, except for the [\FeII]~1.644\,\um\ line. The mean and standard deviation of the peak velocities of the first and second epochs are $-$81~$\pm$~2\,km\,s$^{-1}$ and $-$110~$\pm$~59\,km\,s$^{-1}$, respectively. It is hard to compare these lines quantitatively because of the lower S/N and spectral resolution of the second epoch. However, these line variations suggest that the physical properties of jets changed between the two epochs. 

\begin{figure}
    \centering
    \includegraphics[width=\columnwidth]{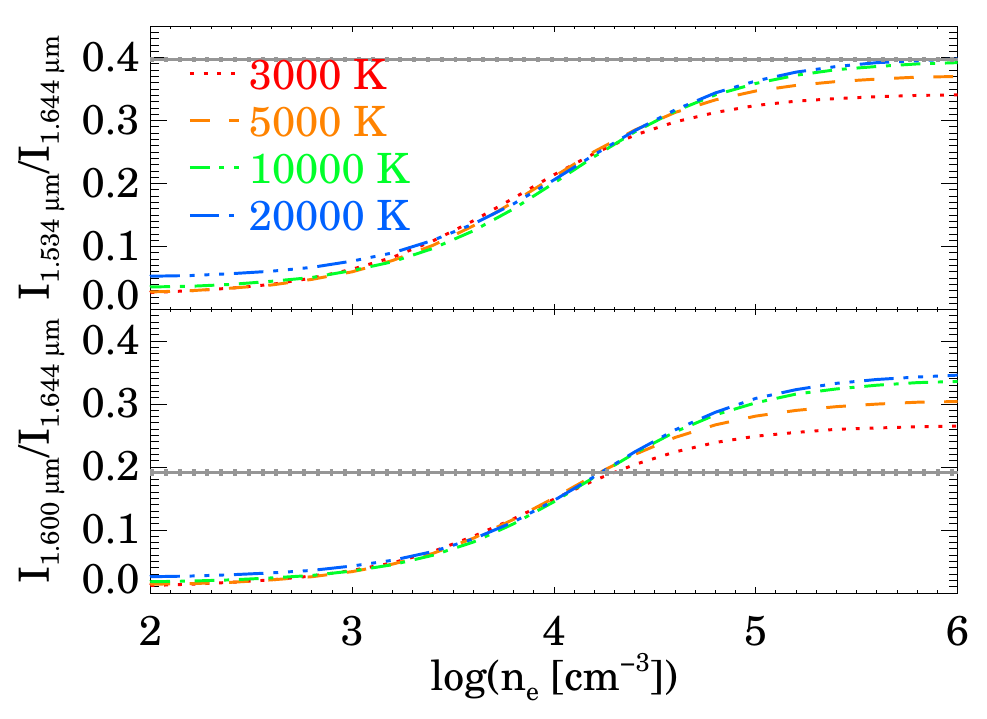}
    \caption{Observed [\FeII] line ratios for Gaia19fct. Dotted red, dashed orange, dashed-dotted green, and dashed-triple dotted blue lines present models with $T_e$=3000, 5000, 10000, and 20000\,K, respectively. Gray horizontal lines and ticks represent the observed line ratios and uncertainties.
    \label{fig:feii_ratio}}
\end{figure}

The electron density ($n_{e}$) and electron temperature ($T_{e}$) of the jet/outflow region can be estimated using the [\FeII] line ratios. To estimate the physical properties of the jet/outflow region, we only used the reliable first epoch IGRINS spectrum, which has a higher S/N. We used a line ratio (log([\FeII] 1.644/1.534) = 0.40) of \citet{nisini2002} and compared it with their Figure~8. The calculated ratio implies $n_{e}$ and $T_{e}$ are higher than 10$^5$\,cm$^{-3}$ and 15000\,K, respectively.
Then, we used the CHIANTI database\footnote{\url{http://chiantidatabase.org}} to calculate the model line ratios as done in \citet{kospal2020}. \autoref{fig:feii_ratio} shows the models and observed [\FeII] line ratios. 
The upper panel ([\FeII] 1.534/1.644 =  0.397~$\pm$~0.002) implies that $n_{e}$ and $T_{e}$ of Gaia19fct are higher than 10$^5$\,cm$^{-3}$ and 10000\,K, respectively, while the lower panel ([\FeII] 1.600/1.644 = 0.191~$\pm$~0.003) shows $n_{e}$ is between 10$^4$ and 10$^5$\,cm$^{-3}$, but the temperature cannot be constrained using the observed line ratios because the model curves are degenerate.
The obtained [\FeII] line ratios suggest that the jet/outflow region of Gaia19fct has $n_{e}$ and $T_{e}$ higher than 10$^4$\,cm$^{-3}$ and 10000\,K, respectively.

\subsubsection{Metallic lines} \label{sec:metallic_lines}
\begin{figure*}
    \centering
    \includegraphics[width=\textwidth]{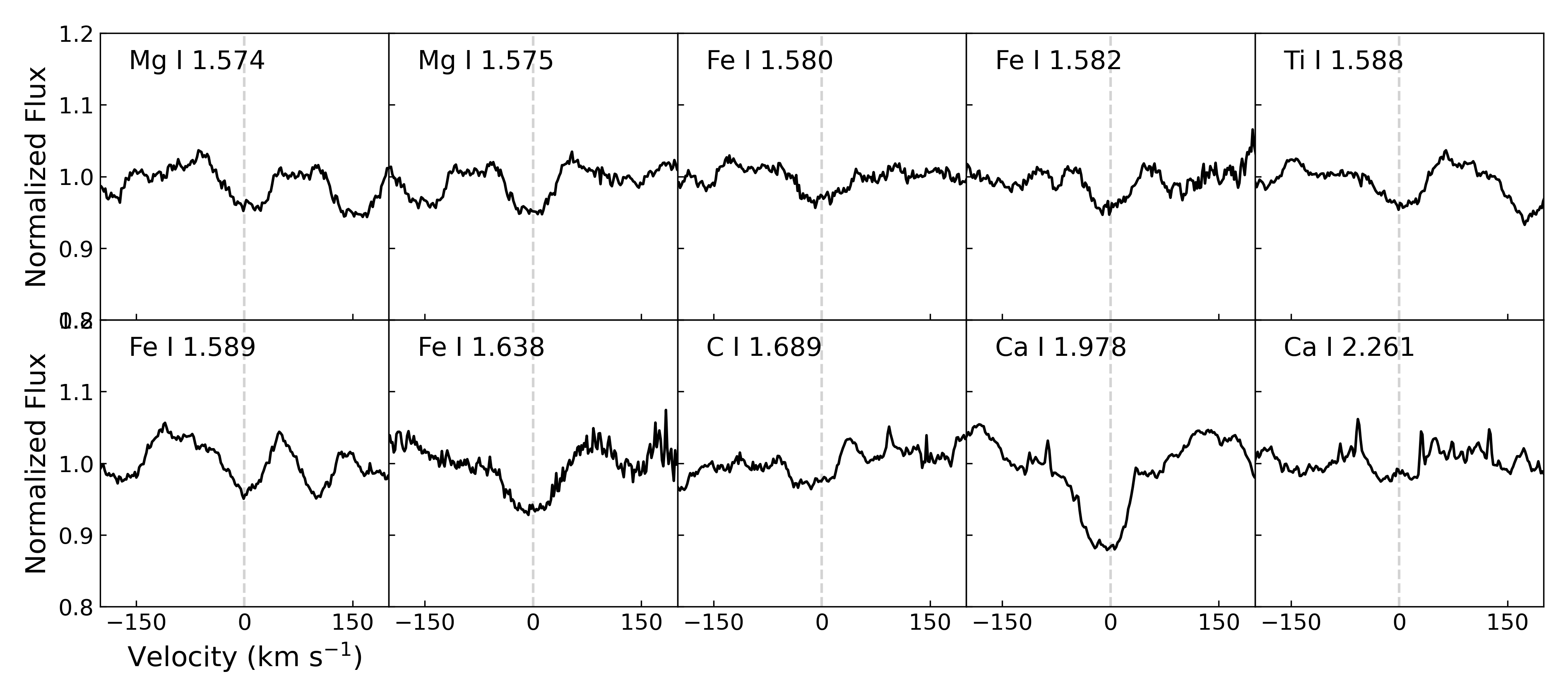}
    \caption{Detected metallic absorption lines in 2020 November. 
    \label{fig:metallic}
    }
\end{figure*}

In addition to emission lines related to the accretion or jet/outflow, several metallic absorption lines are observed in 2020 November. \autoref{fig:metallic} shows relatively isolated and strong lines.
The atomic metallic lines can form in a rotating disk or a protostellar photosphere. In the case of the FUors, the absorption line profiles can form at a cooler disk photosphere seen in front of a hotter disk midplane, and the line profiles are double-peaked or boxy due to the Keplerian rotation of the disk \citep{hartmann1996, fischer2022}. The broad single-peaked line profiles can form at the rotating protostellar photosphere \citep{gray1992}. 
The observed NIR metallic lines show u-like or boxy profiles, similar to optical lines presented by \citet{hillenbrand2019}, rather than a clear double-peaked profile. In order to investigate the origin of the metallic lines, we fitted these atomic absorption lines with standard stellar spectrum from IGRINS Spectral Library \citep{park2018} by convolving with disk rotational and stellar rotational profiles described in \citet{yoon2021}. 

\begin{figure*}
    \centering
    \includegraphics[width=0.32\textwidth]{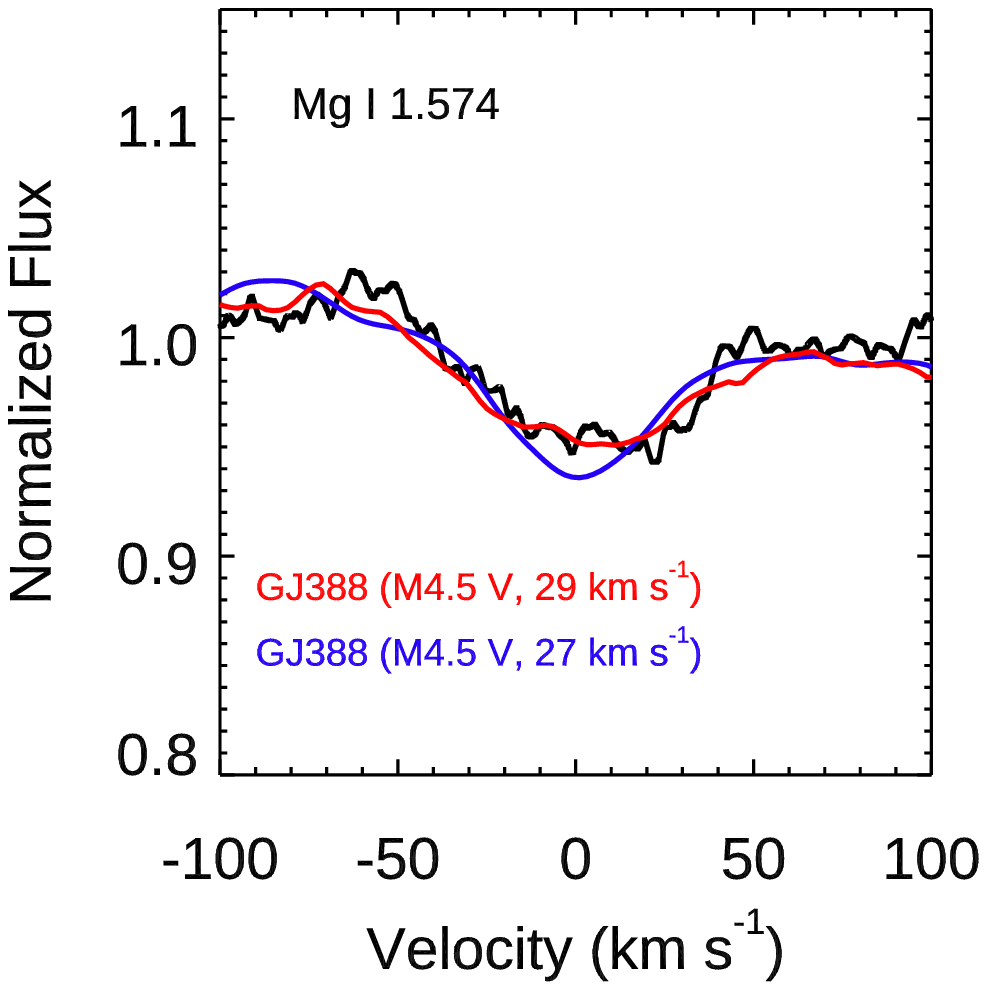} 
    \includegraphics[width=0.32\textwidth]{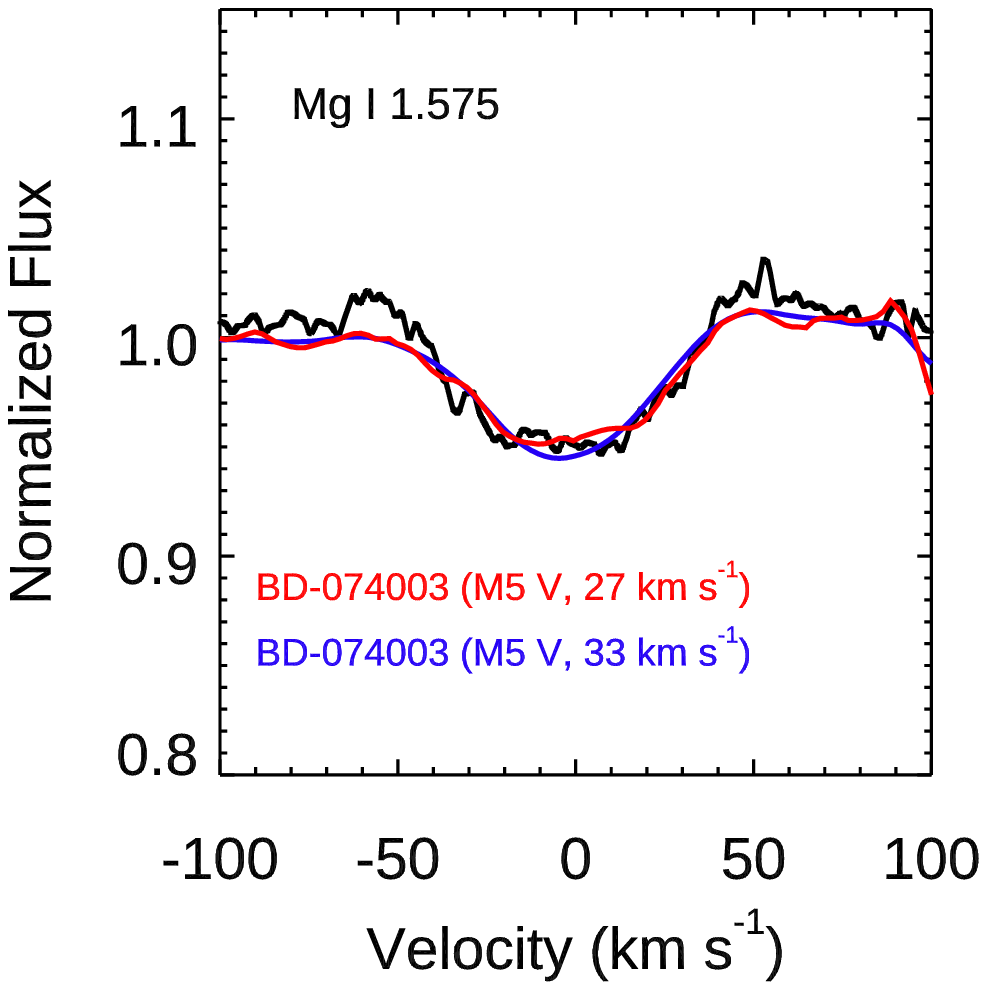}  \includegraphics[width=0.32\textwidth]{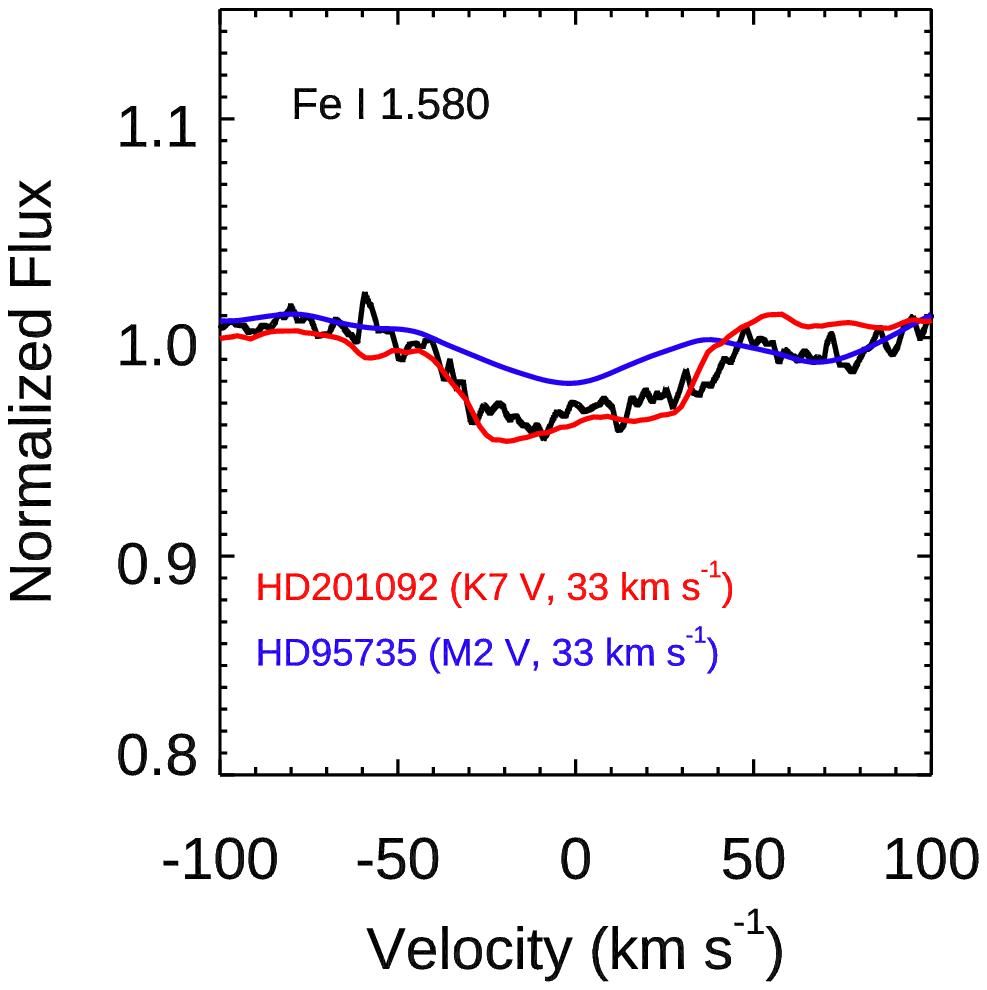}
    \caption{Best fit results of \MgI\ 1.574, \MgI\ 1.575, and \FeI\ 1.580\,\um. Black line shows the error weighted mean spectrum of Gaia19fct. Red and blue lines present disk and stellar rotational profiles, respectively.
    \label{fig:convol}
    }
\end{figure*}

Most of the observed lines are blended with adjacent lines compared to the convolved stellar spectrum, making only three lines available for the fitting. The best fit was found by ${\chi}^2$ minimization, and \autoref{fig:convol} shows the best-fit results for the \MgI\ 1.574, \MgI\ 1.575, and \FeI\ 1.580\,\um\ lines. 
The wavelengths of the three lines are close to each other, therefore, the same temperatures and spectral types are expected within the uncertainties. Due to the small number of lines used and the limited grid of spectral types and luminosity classes for standard stars, we provide the ranges of spectral types and rotational velocities.
The observed NIR lines are well-fitted by a relatively cooler temperature (from K7~V to M5~V) with lower velocities (between 27 and 33\,km s$^{-1}$) compared to the optical lines \citep[F5 to G0 depending on luminosity class and about 70\,km s$^{-1}$;][]{hillenbrand2019}.
Additionally, the boxy profiles are better fitted with disk rotational profiles. This result shows the wavelength-dependent spectral type, typical of FUors, and suggests that these NIR metallic lines are formed at the cooler outer part of the disk in Keplerian rotation.

\subsubsection{Stellar and Accretion Parameters}
The estimation of the stellar parameters for embedded Class~I objects is quite challenging, since they are difficult to observe in the visible. As a consequence, the spectral typing is limited to the NIR regime, where the contamination due to the extinction and the veiling is not negligible. The embedded nature of protostars is also a limitation to the direct estimate of the accretion luminosity because the Balmer jump in the near-UV is unobservable in them. This prevents us from studying stellar and accretion parameters of Class~I YSOs using the same methods as used for Class~II stars based on Balmer jump, absorption lines, and spectral shape \citep[see for example,][for spectral typing in Class~II YSOs]{manara2013}.

We estimated stellar parameters and accretion luminosity of Gaia19fct by modifying a self-consistent method usually used for Class~I and flat YSOs analysis \citep{antoniucci2008, fiorellino2021}.
This method is based on the following assumptions: ($i$) the bolometric luminosity is the sum of the stellar and the accretion luminosity ($L_{\rm bol} = L_\star + L_{\rm acc}$); ($ii$) the absolute bolometric magnitude in $K$\,band is M$_{\rm bol} = \mbox{BC}_K + m_K + 2.5 \log(1 + r_K) - A_K - 5 \log(d/10 \mbox{pc})$; 
($iii$) the empirical relations between the luminosity of {H\,\footnotesize{I}} lines and accretion luminosity found for CTTS are a good approximation for Class~I stars \citep{nisini2005}, in particular the one regarding the Br$\gamma$ line, $\log L_{\rm acc} = a \log L_{Br\gamma} + b$, where $a = 1.19 \pm 0.10$ and $b = 4.02 \pm 0.51$ \citep{alcala2017}. 
All these equations depend on the extinction. 
This method typically takes the measured {H\,\footnotesize{I}} flux, estimation of veiling, distance, and bolometric luminosity as input. Output includes extinction, stellar mass, radius, luminosity, spectral type (resulting from the bolometric correction BC$_K$), and accretion luminosity.

Since this approach is valid for accreting objects until the magnetospheric accretion scenario works \citep{antoniucci2008, fiorellino2021}, we checked the spectrum we used for this analysis corresponds to a quiescent phase. Looking at \autoref{fig:light_curve}, we see that this condition is satisfied only for the NOTCam spectrum observed in 2022, as the two earlier spectra were taken during the fading phase of a brightening event. Moreover, since we need the estimation of the flux, contemporary photometry is also required. For this epoch, we have simultaneous photometric observation. We calculated the flux of the Br$\gamma$ line by using the contemporary photometry in $K$ band, $m_K = 10.476 \pm 0.009$\,mag, obtaining $F_{{\rm Br}\gamma} = (3.89 \pm 0.06) \times 10^{-15}$\,erg\,s$^{-1}$\,cm$^{-2}$. We adopted the distance of 0.92\,kpc (Section~\ref{sec:distance}) and ${L_{\rm bol}}$ of 5.6\,$L_\odot$ as we computed in Section~\ref{sec:classification}.
We assumed ages of 10$^4$\,yr (birthline) and 1\,Myr since the age of this source is unknown. Then, stellar parameters that satisfied equation ($i$) and located near the birthline and the 1\,Myr line were found.
Since we were not able to compute the veiling for this target, we set it as a free parameter, varying its value from 0 to 20, with steps of 0.1, looking for the set of parameters in agreement with our extinction estimate, i.e., $A_V=8\pm1$\,mag (Section~\ref{sec:extinction}). 
As a result, we obtained the veiling ($r_K$), stellar mass (\Mstar), radius (\Rstar), and luminosity (\Lstar) which satisfy equations ({\it i}), ({\it ii}), and ({\it iii}) using $F_{{\rm Br}\gamma}$, $\Lbol$, and $A_V$, we measured. The obtained values are listed in Table~\ref{tbl_param}.

From these parameters, we computed the mass accretion rate ($\dot{M}_{\rm acc}$) using:
\begin{equation} \label{eq_Macc}
    \dot{M}_{\rm acc}=\left(1- \frac{R_{*}} {R_{\rm in}}\right)^{-1}~\frac{L_{\rm acc} R_{*}} {G M_{*}},
\end{equation}
where $R_{\rm in}$ is the disk inner radius and assumed to be $5\,R_{*}$~\citep{hartmann1998}.
The calculated $\dot{M}_{\rm acc}$ for the birthline and 1\,Myr is $(2.1 - 2.6)\times10^{-7}$\,$M_{\odot}$ yr$^{-1}$ and $(1.2 - 1.5)\times10^{-7}$\,$M_{\odot}$ yr$^{-1}$, respectively, which are lower than typical of FUors, while similar to EXors or CTTS \citep[][and references therein]{fischer2022}.
In addition, the obtained \Macc\ is compatible with Class~I sources in the NGC~1333 cluster \citep[see Figure~12 from][]{fiorellino2021}, consistent with the evolutionary stage of Gaia19fct. One should keep in mind that our method is valid assuming magnetospheric accretion, resulting in the calculated mass accretion rate being a lower limit in the case of the eruptive stars. Additionally, veiling was found as a free parameter; therefore, a detailed study about veiling is needed for more accurate estimate of the accretion rate.  
The average errors on the accretion luminosity, stellar radius, and mass are 0.4, 0.6, and 0.1\,dex, respectively, which result in a cumulative error for the mass accretion rate of 0.8\,dex.

\begin{deluxetable}{lcc}
\tablecaption{Stellar and Accretion Parameters \label{tbl_param}}
\tablehead{\colhead{Parameter} & \colhead{Birthline (10$^4$\,yr)}  & \colhead{1\,Myr} }
\startdata
$r_K$ & 0.7 $-$ 1.1 & 1.0 $-$ 1.5 \\
Spectral Type &  M1 & K7 \\
\Teff\ (K) & 3631 & 3981 \\
\Lstar\ ($L_{\odot}$) & 5.16 $\pm$ 0.41 & 5.05 $\pm$ 0.12  \\
\Mstar\ ($M_{\odot}$) & 0.44 $\pm$ 0.01 & 0.70 $\pm$ 0.01  \\
\Rstar\ ($R_{\odot}$) & 5.26 $\pm$ 0.21 & 4.67 $\pm$ 0.19  \\
\Lacc\  ($L_{\odot}$) & 0.50 $\pm$ 0.06 & 0.50 $\pm$ 0.06 \\
$\dot{M}_{\rm acc}$\ (10$^{-7}$ $M_{\odot}$\,yr$^{-1}$) & 2.10 $-$ 2.63 & 1.17 $-$ 1.47 \\
\enddata
\end{deluxetable}

\subsubsection{CO modeling} \label{sec:co}
The CO overtone feature in Gaia19fct appears to be a superposition of absorption and emission components, unlike what is typical in other young eruptive stars or T\,Tauri stars. The high spectral resolution of IGRINS allowed us to resolve the individual rotational lines of this feature. To determine the velocities of each component, we first normalized each rotational line of the $v=2-0$ transition and took their mean. Then we fitted Gaussian functions to this mean line profile. We found that the profile is best fitted with the sum of an emission component at 7.0\,km\,s$^{-1}$ and two absorption components at $-$29.8\,km\,s$^{-1}$ and $-$60.5\,km\,s$^{-1}$. This solution matches the mean $v=3-1$ line profile as well, but we did not do a separate fit for it because of its lower S/N. The velocities given here are relative to the systemic velocity of $V_{LSR}$ = 13.25~km\,s$^{-1}$ of Gaia19fct. 

\begin{figure*}
    \centering
    \includegraphics[width=\textwidth]{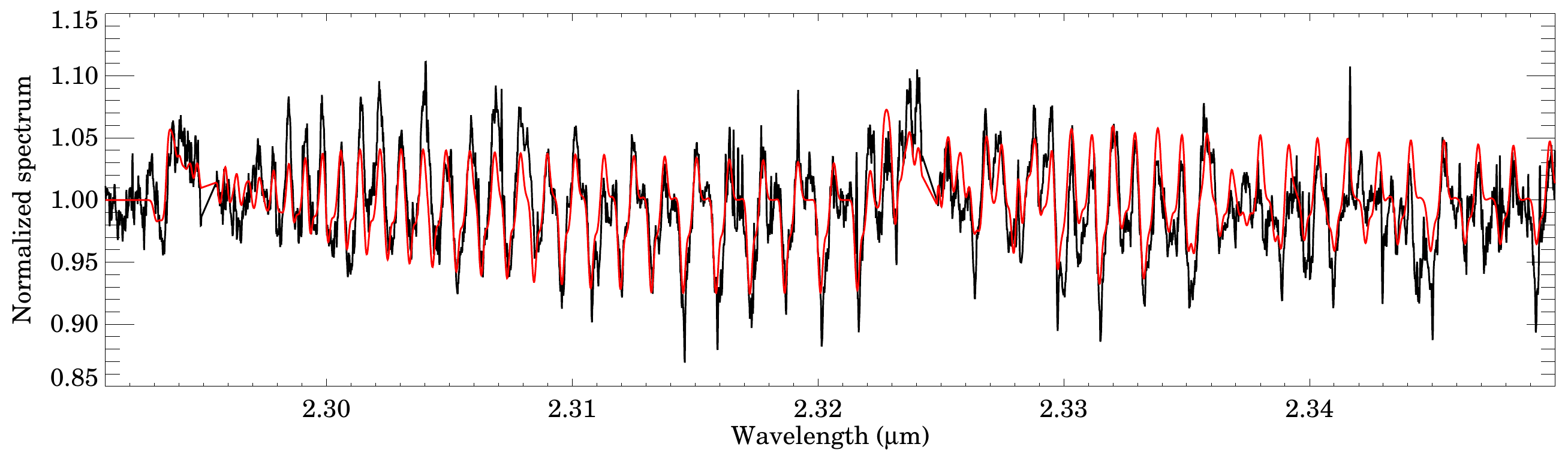}
    \includegraphics[width=\textwidth]{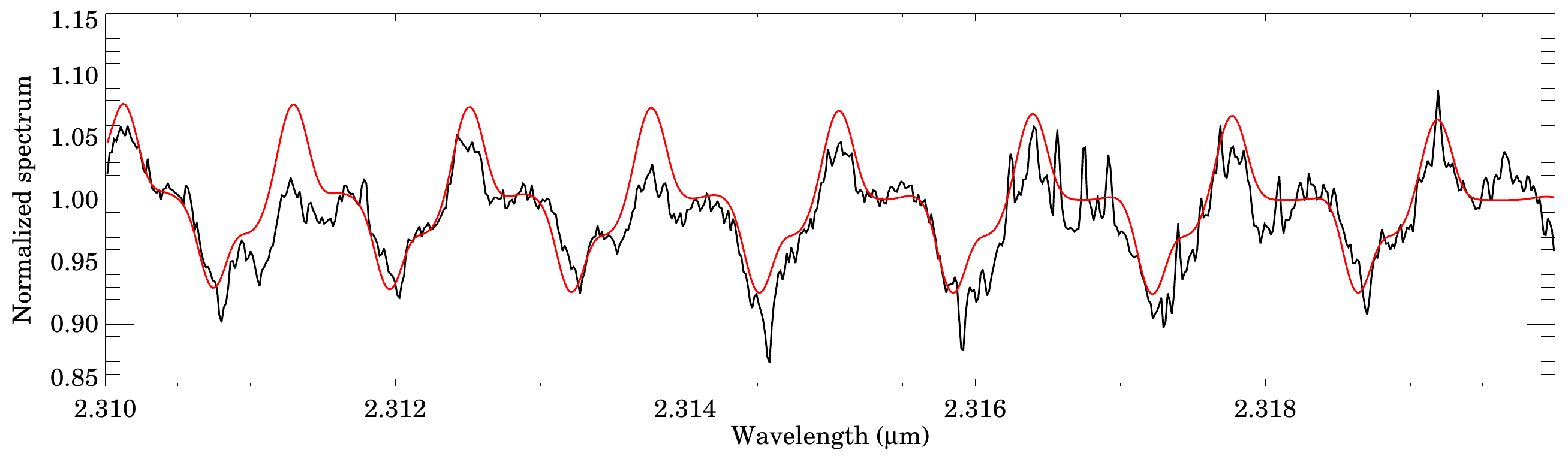}
    \caption{CO overtone features of Gaia19fct as observed on 2020 November 23. Black line shows the spectrum of Gaia19fct and red line represents the  best fit result. The upper panel shows the full $v=2-0$ and $v=3-1$ series of lines, while the lower panel is a zoom-in for some of the individual rotational transitions, clearly displaying the emission and two absorption components that make up the spectrum of Gaia19fct.}
    \label{fig:co}
\end{figure*}

The varying ratio of the emission and absorption components suggested that the temperature and column density of the emitting and absorbing material are different. To quantify this, we used a simple isothermal slab model to calculate the expected CO overtone feature spectrum for a grid of different excitation temperatures ($T$) and CO column densities ($N_{\rm CO}$), following the approach described in \citet{kospal2011} and \citet{park2021_v899mon}. 
For this analysis, we fitted both IGRINS spectra separately to provide ranges of the obtained physical parameters. 
The velocities of the three components were fixed to the values determined above. For simplicity, we also prescribed that $T$ and $N_{\rm CO}$ are the same for the two absorption components. Thus, we had four free parameters in our fitting: $T$(emission), $N_{\rm CO}$(emission), $T$(absorption), and $N_{\rm CO}$(absorption). We found the minimal $\chi^2$ with $T$(emission) = 4100\,K,
$N_{\rm CO}$(emission) = $3.5\times10^{21}$\,cm$^{-2}$,
$T$(absorption) = 1200\,K, and
$N_{\rm CO}$(absorption) = $2.2\times10^{21}$\,cm$^{-2}$ for the spectrum taken on 2020 November 19, and
$T$(emission) = 3200\,K,
$N_{\rm CO}$(emission) = $4.0\times10^{21}$\,cm$^{-2}$,
$T$(absorption) = 1300\,K, and
$N_{\rm CO}$(absorption) = $2.8\times10^{21}$\,cm$^{-2}$ for the spectrum taken on 2020 November 23. The latter spectrum and the corresponding model is shown in \autoref{fig:co}.

Because the two IGRINS spectra taken with four days apart agree well within the measurement uncertainties, the difference in the fitted parameters for the two spectra can be regarded as a confidence interval. Therefore, we conclude that the CO-emitting material is hotter, in the 3200-4100\,K range, while the CO-absorbing material is cooler, in the 1200-1300\,K range. Similarly, the CO-emitting material is optically thicker, with column densities in the $(3.5 - 4.0)\times10^{21}$\,cm$^{-2}$, while the CO-absorbing material is optically thinner, in the $(2.2 - 2.8)\times10^{21}$\,cm$^{-2}$ range.

Composite CO spectra, with the superposition of absorption and emission components at various velocities, are rarely seen in any kind of astronomical object. In the following, we discuss a few such examples we found in the literature, which may help is interpret what we see in Gaia19fct. \citet{geballe2007} observed highly  variable composite CO overtone feature in the luminous red nova V838 Mon, where the absorption is interpreted as partly coming from the stellar photosphere, partly from high-velocity gas ejected by the outburst. \citet{gorlova2006} detected a composite structure for the CO overtone band in the pulsating yellow hypergiant $\rho$ Cas, with both the emission and the absorption components variable. Here, the different components originate in different atmospheric layers. \citet{harrison2016} observed the cataclysmic variable WZ Sge and found that its CO overtone feature has a central dip of absorption superimposed on emission. In this case, the absorption is coming from a substellar L dwarf companion, while the emission comes from the accretion disk of white dwarf primary. \citet{brittain2005ApJ} observed the fundamental CO lines towards the heavily embedded young star HL Tau and found narrow central absorption superimposed on broad emission with the same central velocity. The broad CO emission originates from the hot ($T\sim1500$\,K) inner disk, while the narrow absorption is caused by a large column of cold ($T\sim100$\,K) material.

None of the models or explanations found in the literature can be directly applied to Gaia19fct. Both the emission and absorption components here are relatively narrow (in the $8 - 12$\,km\,s$^{-1}$ range while the instrumental line broadening is 7\,km\,s$^{-1}$). Therefore, it is unlikely that the emission is coming from the hot inner accretion disk, as usual with YSOs with high accretion rates, but points to the possibility of a slow, dense stellar wind with high mass loss rate instead \citep{carr1989}. The blueshifted absorption components suggest multiple, expanding circumstellar shells, possibly launched by the stellar or disk wind.

\section{Discussion} \label{sec:discussion} 
Gaia19fct has been studied in several works \citep{miller2015, hillenbrand2019, giannini2022} after its discovery in 2015 \citep{miller2015}. Since 2015, at least five burst events have occurred, and each of these events shows a different amplitude, duration, and speed of brightness variation ($\Delta$\,mag/$\Delta$\,t, Section~\ref{sec:light_curve}), indicating that similar to the classical EXor V1118~Ori \citep{giannini2020}, these bursts are not the periodic repetition of the same event. This is supported by the fact that the light curve of Gaia19fct do not present any clear periodicity on both large and small time scales (Section~\ref{sec:small_scale}).

\citet{sewilo2019} fitted the SED with a star and disk-only system and provided the best-fit results: \Teff $\sim$ 9620\,K, \Lstar$\sim$111\,$L_{\odot}$, \Mstar$\sim$3\,$M_{\odot}$, \Rstar$\sim$3.8\,$R_{\odot}$, and Age$\sim$2.13\,Myr. These stellar parameters are not only higher than most of the FUors studied by \citet{gramajo2014}, including an intermediate-mass eruptive star Z~CMa, but also higher than Class~I sources studied by \citet{fiorellino2021}.
In addition, according to our classification (Section~\ref{sec:classification}) and previous studies \citep{miller2015, fischer2016}, Gaia19fct is classified as Class~I rather than a star and disk-only system, which is more evolved. Therefore, we revisited the stellar parameters by modifying the self-consistent method described in \citet{fiorellino2021} and provided the results in Table~\ref{tbl_param}. The obtained parameters are in agreement with FUors \citep{gramajo2014} and Class~I sources \citep{fiorellino2021}.

\subsection{Comparison with FUors}
In our NIR spectra, we observed boxy or u-shape absorption line profiles typically found in the Keplerian rotating disk of FUors. We fitted these lines by convolving the spectra of standard stars with disk rotational profiles and found the best-fit results (Section~\ref{sec:metallic_lines}). The estimated rotational velocity is smaller than those of optical lines \citep{hillenbrand2019}, and the best-fit spectral type is also later than the optical spectrum \citep{hillenbrand2019}. These results suggest that Gaia19fct has a Keplerian rotating disk. If this target has a Keplerian disk, the temperature and rotational velocity decrease with the radial extension of the disk \citep[][and references therein]{fischer2022}. 
With the obtained rotational velocities, we estimated the location of the disk where these observed lines are formed by assuming an inclination of 90$^{\circ}$ and a stellar mass of 0.7\,$M_{\odot}$. We adopted the stellar mass obtained by assuming 1\,Myr, which is more reasonable to the Class~I \citep{fiorellino2021}, based on our classification (Section~\ref{sec:classification}). 
For the optical lines, we adopted the rotational velocity (70\,km\,s$^{-1}$) and spectral type (F0-G0 supergiants to dwarfs) from \citet{hillenbrand2019}. Then, we assumed the temperature of 6359\,K as the median spectral type of F5~V star HD~87141 \citep[6359\,K;][]{prugniel2011}. For the NIR lines, we used the mean velocity of 30\,km\,s$^{-1}$ from our estimation (Section~\ref{sec:metallic_lines}) and assumed the mean temperature of 3650\,K between K7~V star HD~201092 \citep[3911\,K;][]{prugniel2011} and M5~V star BD-07~4003 \citep[3209\,K;][]{santos2013}. 

\begin{figure}
    \centering
    \includegraphics[width=\columnwidth]{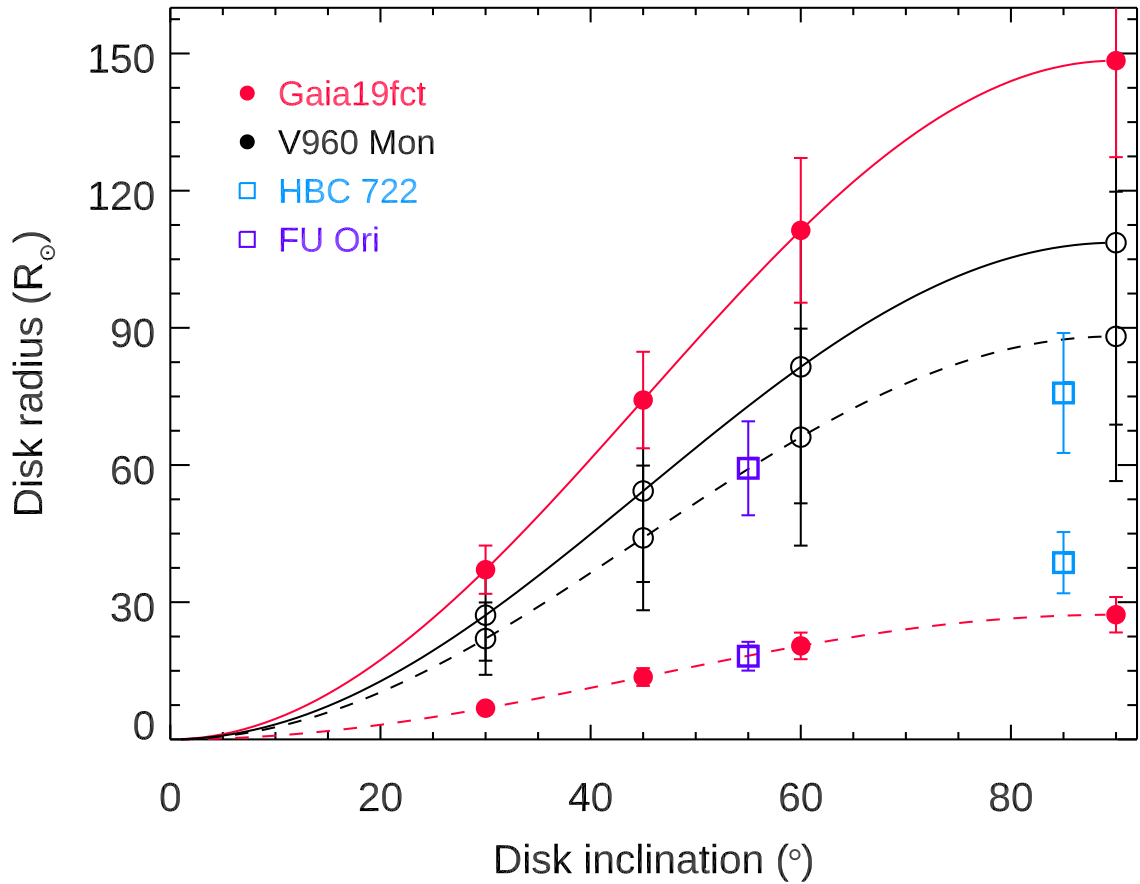}
    \caption{Calculated disk radius as a function of disk inclination. Different colors indicate different targets. Dashed and solid lines represent the radius calculated with the best-fit rotational velocity of optical and NIR lines (Table~\ref{tbl:comp_disk}), respectively. The circle symbols present the radius calculated with inclinations of 30$^{\circ}$, 45$^{\circ}$, 60$^{\circ}$, and 90$^{\circ}$.
    \label{fig:comp_disk}
    }
\end{figure}

If we assume the disk inclination of 90$^{\circ}$ ($V_{max}=vsini$) and stellar mass of 0.7\,${M_{\odot}}$, the observed optical and NIR boxy or u-shape absorption lines trace $27 \pm 4$\,$R_{\odot}$ and $148 \pm 21$\,$R_{\odot}$ of the disk, and the temperature of the disk decreases from 6359\,K at $27 \pm 4$\,\Rsun to 3650\,K at $148 \pm 21$\,$R_{\odot}$. Spectral features at shorter wavelengths trace the hotter inner part of the disk with higher rotational velocity. In comparison, the features at longer wavelengths trace the cooler outer part of the disk with lower rotational velocity. This result is consistent with the Keplerian rotating disk profile of FUors, including HBC~722 \citep{lee2015} and V960~Mon \citep{park2020}. We also estimated the location of the disk traced by various spectral features for the prototype FU~Ori by using the literature values listed in Table~\ref{tbl:comp_disk}. The estimated radial distances traced by optical and NIR lines are about $18 \pm 3$\,$R_{\odot}$ and $59 \pm 10$\,$R_{\odot}$, respectively. 
Since the inclination of Gaia19fct is unknown, we calculated the disk radius by varying disk inclination, as shown in \autoref{fig:comp_disk}. Depending on the inclination, the observed lines trace different radii. If the disk inclination is higher than 45$^{\circ}$, the observed NIR lines of Gaia19fct trace a larger disk radius than those of three FUors. In contrast, optical lines trace a  smaller disk radius than V960~Mon and HBC~722 but trace a smaller or larger disk radius than FU~Ori.
If the disk inclination is smaller than 30$^{\circ}$, optical and NIR lines trace a smaller disk radius than HBC~722 and FU~Ori and still trace a smaller disk radius than V960 Mon in optical, but trace a larger disk radius in NIR.
Since the observed boxy or u-shape line profiles do not present clear double-peaked profiles, the disk inclination might not be high. To constrain the disk radius traced by observed lines precisely, further study about disk inclination is needed.

\begin{deluxetable}{cccccccc}
\tabletypesize{\scriptsize}
\tablecaption{Comparison with FUors} \label{tbl:comp_disk}
\tablewidth{0pt}
\tablehead{\colhead{Wavelength} & \colhead{Mass} & \colhead{Target} & \colhead{\Vmax} & \colhead{Spectral Type} & \colhead{Temperature} & \colhead{Radius$^\dagger$} & \colhead{References} \\
\colhead{} & \colhead{(\Msun)} & \colhead{} & \colhead{(\kms) } & \colhead{} & \colhead{(K)} & \colhead{(\Rsun)} & \colhead{}}
\startdata
Optical     & 0.70 $\pm$ 0.01           & Gaia19fct             & 70          & F0 - G0 I-V        & 6359$^{a}$  & 27$\pm$4 & 1, 2 \\
            & 0.75 $\pm$ 0.25           & V960~Mon              & 40.3$\pm$3.8 & G2 II-III / G5 III & 5308 / 5013  & 88$\pm$32 & 3, 4, 5, 6 \\
	  	    & 0.8 $\sim$ 1.0.           & HBC~722            	& 70           & G5 II 	            & 5090        		 & 39$\pm$7 & 7, 8, 9, 10 \\
	  	    & 0.60                      & FU~Ori                & 65$\pm$5     & F8 - G4            & 6420               & 18$\pm$3 & 11, 12 \\
NIR	        & 0.70 $\pm$ 0.01 		    & Gaia19fct             & 30$\pm$3     & K7 V - M5 V 	    & 3650$^{b}$   		 & 148$\pm$21 & \\
            & 0.75 $\pm$ 0.25 		    & V960~Mon              & 36.3$\pm$3.9 & K1 III 	        & 4634   		 & 109$\pm$40 & 3, 4, 13 \\
		    & 0.8 $\sim$ 1.0   		    & HBC~722 			    & 50           & K5 Iab 	        & 3055    		 & 76$\pm$13 & 7, 8, 9, 14 \\
            & 0.60                      & FU~Ori                & 36$\pm$3     & M                  &                   & 59$\pm$10 & 10, 12\\    
\enddata
\tablenotetext{\dagger}{~Radius was calculated by using the maximum projected velocity (\Vmax).}
\tablenotetext{a}{~We assumed the temperature of F5~V star (HD~87141) because \citet{hillenbrand2019} suggested the spectral type between F0 to G0 depending on luminosity classes.}
\tablenotetext{b}{~Mean temperature of K7~V star (HD~201092) and M5~V star (BD-07~4003) was used.} 
\tablerefs{
(1) \citet{hillenbrand2019}; (2) \citet{prugniel2011}; (3) \citet{kospal2015}; (4) \citet{park2020};
(5) \citet{liu2014}; (6) \citep{park2018};
(7) \citet{kospal2016}; (8) \citet{gramajo2014}; (9) \citet{lee2015};
(10) \citet{kovtyukh2007}; (11) \citet{perez2020}; (12) \citet{zhu2009};
(13) \citet{wu2011}; (14) \citet{bakos1971}
}
\end{deluxetable}

\subsection{Comparison with EXors}
Weak CO overtone bandhead features superimposed with absorption profiles were observed in 2020 November. The CO overtone emission features are typically found in EXors and EXor-like objects \citep{ kospal2011, hodapp2019, hodapp2020, park2021_v899mon, csm2022}. To study the physical properties (\Tex\ and $N_{\rm CO}$) where these CO features are formed, we fitted a simple slab model for the emission and absorption components separately and found the best fit results (Section~\ref{sec:co}). 
We compared the fitting result of the emission component, which is formed in the hot inner accretion disk, with the prototype EXors EX~Lup and EXor-like object V899~Mon. The CO excitation temperature of Gaia19fct is higher ($3200 - 4100$\,K) than those of EX~Lup \citep[2500\,K;][]{kospal2011} and V899~Mon \citep[$2482 \pm 326$\,K;][]{park2021_v899mon}, suggesting that the inner disk, where CO features are formed, of Gaia19fct is hotter than the two other objects.

\section{Conclusions} \label{sec:conclusion}
We have conducted optical and NIR photometric and NIR spectroscopic observations of the young eruptive star Gaia19fct since 2016 September. We analyzed our observations along with public domain data to study the physical properties of Gaia19fct. From our analysis, our major conclusions are as follows.

1. Gaia19fct has been undergoing brightening events at least five times since its discovery in 2015 and keeps changing its brightness. The moderate amplitudes ($\Delta r = 2.5 - 5$\,mag) and short time scales ($<$ 1\,yr) of bursts are similar to EXors.

2. Overall gray variability from 2019 to 2022 suggests that a mechanism other than extinction change might cause the brightness variation. 

3. We classified Gaia19fct using several methods: \Tbol, $\alpha$, and IR colors. \Tbol\ and $\alpha$ suggest that the evolutionary stage of Gaia19fct is Class~I, while the IR colors show Class~I and flat-spectrum. Based on our data, we suggest that Gaia19fct is in the Class~I stage.

4. Our NIR spectra show both absorption and emission lines, similar to the characteristics of FUors and EXors, and the observed lines varied with time.

5. Several atomic metal lines are observed in absorption, and well-matched with K7 to M5 dwarfs convolved with disk rotational profiles of about 30\,km s$^{-1}$. This result agrees with the wavelength-dependent spectral type of FUors compared with those of the optical spectrum \citep{hillenbrand2019}.

6. We calculated stellar and accretion parameters of Gaia19fct, finding that the stellar parameters are similar to low-mass, late-type stars, and the resulting mass accretion rate is more similar to EXors than to FUors.

7. The observed CO features have a composite structure of emission and absorption components, and our fitting result suggests that the emission is formed close to the star and the absorption is formed by expanding shells.

8. Our results show that Gaia19fct displays photometric and spectroscopic characteristics of both FUors and EXors, but it shows more similarity to EXors.
The mixture of FUor and EXor properties is expected to provide important insights into our understanding of the accretion process in eruptive young stars. Therefore, further photometric and spectroscopic monitoring of Gaia19fct is needed.


\bibliographystyle{aasjournal}
\bibliography{gaia19fct}

\begin{acknowledgments}
This work used the Immersion Grating Infrared Spectrometer (IGRINS) that was developed under a collaboration between the University of Texas at Austin and the Korea Astronomy and Space Science Institute (KASI) with the financial support of the Mt. Cuba Astronomical Foundation, of the US National Science Foundation under grants AST-1229522 and AST-1702267, of the McDonald Observatory of the University of Texas at Austin, of the Korean GMT Project of KASI, and Gemini Observatory.
This work was supported by K-GMT Science Program (PID: GS-2020B-Q-218) of Korea Astronomy and Space Science Institute (KASI).

This project has received funding from the European Research Council (ERC) under the European Union's Horizon 2020 research and innovation programme under grant agreement No 716155 (SACCRED), and from the ``Transient Astrophysical Objects'' GINOP 2.3.2-15-2016-00033 project of the National Research, Development and Innovation Office (NKFIH), Hungary, funded by the European Union.
We acknowledge support from the ESA PRODEX contract nr. 4000132054.
Zs.N., L.K., and K.V. acknowledge the support by the J\'anos Bolyai Research Scholarship of the Hungarian Academy of Sciences. K.V. is supported by the Bolyai+ grant UNKP-22-5-ELTE-1093. 
This project has been supported by the K-131508 grant of the Hungarian National Research, Development and Innovation Office (NKFIH) and the \'Elvonal grant KKP-143986. Authors acknowledge the financial support of the Austrian-Hungarian Action Foundation (101.u13, 104.u2).  
L.K. acknowledges the financial support of the Hungarian National Research, Development and Innovation
Office grant NKFIH PD-134784.

This publication makes use of data products from the Wide-field Infrared Survey Explorer, which is a joint project of the University of California, Los Angeles, and the Jet Propulsion Laboratory/California Institute of Technology, funded by the National Aeronautics and Space Administration. This publication also makes use of data products from NEOWISE, which is a project of the Jet Propulsion Laboratory/California Institute of Technology, funded by the Planetary Science Division of the National Aeronautics and Space Administration.

We would like to thank all contributors of observational data for their efforts towards the success of the HOYS project.

Based on observations made with the Nordic Optical Telescope, owned in collaboration by the University of Turku and Aarhus University, and operated jointly by Aarhus University, the University of Turku and the University of Oslo, representing Denmark, Finland and Norway, the University of Iceland and Stockholm University at the Observatorio del Roque de los Muchachos, La Palma, Spain, of the Instituto de Astrofisica de Canarias.

This project has received funding from the European Union's Horizon 2020 research and innovation programme under grant agreement No 101004719 (OPTICON-RadioNet Pilot). This material reflects only the authors views and the Commission is not liable for any use that may be made of the information contained therein.
\end{acknowledgments}



\end{document}